\documentclass[showpacs,preprintnumbers,amsmath,amssymb,floatfix,nofootinbib]{revtex4}

\usepackage{amsmath,amsfonts,bbm,euscript,hyperref}
\usepackage{graphicx}
\usepackage{dcolumn}
\usepackage{bm}
\usepackage{epsfig}
\epsfclipon

\newcommand{\nco}{\newcommand}
\newcommand{\be}{\begin{eqnarray}}
\newcommand{\ee}{\end{eqnarray}}
\newcommand{\al}{\alpha'}
\def\baray{\begin{eqnarray}}
\def\earay{\end{eqnarray}}

\nco{\lra}{\leftrightarrow}

\nco{\sss}{\scriptscriptstyle} \nco{\dphi}{\varphi}
\nco{\lsim}{\mbox{\raisebox{-.6ex}{~$\stackrel{<}{\sim}$~}}}
\nco{\gsim}{\mbox{\raisebox{-.6ex}{~$\stackrel{>}{\sim}$~}}}

\def\IK{\relax{\rm I\kern-.20em K}}
\def\IM{\relax{\rm I\kern-.20em M}}

\def\lsim{\mbox{\raisebox{-.6ex}{~$\stackrel{<}{\sim}$~}}}
\def\gsim{\mbox{\raisebox{-.6ex}{~$\stackrel{>}{\sim}$~}}}
\def\sss{\scriptscriptstyle}

\def\Mpl{M_p}

\def\grad{\vec{\nabla}}

\def\Vsr{V_{\mathrm{sr}}}

\newcommand{\Bb}{B}
\newcommand{\Cc}{C}

\newcommand{\V}{\mathcal{V}}

\begin{document}

\preprint{UMN-TH-3017/11}

\title{Gauge Field Production in Axion Inflation: Consequences for Monodromy, non-Gaussianity in the CMB, and Gravitational Waves at Interferometers}

\author{Neil Barnaby$^1$, Enrico Pajer$^2$, Marco Peloso$^1$}

\affiliation{
$^1$ School of Physics and Astronomy,
University of Minnesota, Minneapolis, 55455 (USA)\\
$^2$ Department of Physics, Princeton University, Princeton, NJ 08544 (USA)\\
}

\date{\today}

\begin{abstract} 
Models of inflation based on axions, which owe their popularity to the robustness against UV corrections, have also a very distinct class of signatures. 
The relevant interactions of the axion are a non-perturbative oscillating contribution to the potential and a shift-symmetric coupling to gauge fields. We review how these couplings affect the cosmological perturbations via a  unified study based on the in-in formalism.  We then note that, when the inflaton  coupling to gauge fields is high enough to lead to interesting observational results, the backreaction of the produced gauge quanta on the inflaton dynamics becomes relevant during the final stage of inflation, and  prolongs  its duration  by about 10 e-foldings.  We extend existing results on gravity wave production in these models  to account for this  late inflationary phase. 
The strong backreaction phase results in an enhancement of the gravity wave signal at the interferometer scales. As a consequence, the signal is potentially observable  at  Advanced LIGO/VIRGO for the most natural duration of inflation in such models.  Finally, we explicitly compute the axion couplings to gauge fields in string theory construction of axion monodromy inflation and identify cases where they can trigger interesting phenomenological effects.
\end{abstract}

\maketitle

\begin{widetext}
\tableofcontents\vspace{5mm}
\end{widetext}

\section{Introduction} 

\subsection{Motivation}


The next few years will bring a large amount of new information about the physics of the very early universe.  Cosmic Microwave Background (CMB) and Large Scale Structure (LSS) probes will measure the primordial cosmological fluctuations with increasing accuracy, and over a widening range of scales.  This increasing precision will allow us to strongly constrain (or perhaps measure) non-Gaussian statistics of the scalar curvature fluctuations.  At the same time, gravitational waves (tensor perturbations) will also be probed with substantial improvements in accuracy, by CMB experiments and also -- at much smaller scales -- with gravitational interferometry.

It is often claimed that the simplest and most natural microscopic models of inflation lead to a rather minimal set of observable predictions for future missions.  According to the standard lore:
\begin{enumerate}
 \item Tensor fluctuations (gravitational waves) are generated entirely by quantum vacuum fluctuations and their spectrum  is simply controlled by the Hubble parameter during inflation. One can realistically hope to characterize them  by just two numbers: the Hubble rate when some pivot scale leaves the horizon,  and the tensor spectral index, $n_T$.  (The spectrum of tensor vacuum fluctuations is detectable only for sufficiently high scale inflation, which may be unnatural without invoking symmetries.)
This signal is too week to be observed in any of the forthcoming gravity waves interferometers \cite{Smith:2005mm}.
 \item non-Gaussianity arises entirely due to the (weak) self-interactions of the inflaton field and is undetectably small: $f_{NL} \sim \mathcal{O}(\epsilon)$ where $\epsilon=-\frac{\dot{H}}{H^2}$ is a slow roll parameter and $f_{NL}$ characterizes the size of the bispectrum \cite{riotto,maldacena,seerylidsey}.  It is often claimed that a detection of primordial non-Gaussianity would be a ``smoking gun'' signature of some non-standard physics in the early universe.\footnote{For example: small sound speed \cite{small_sound}, higher derivatives \cite{nonlocal}, special initial conditions \cite{nonBD1,nonBD2,nonBD}, potentials with sharp features \cite{chen1,chen2}, dissipative effects \cite{trapped,senatore_new}, turning trajectories \cite{turnNG}, post-inflationary effects (such as preheating \cite{preheatNG,preheatNG2}), etc.}
\end{enumerate}
We argue that, even in very natural models of inflation, the situation may be much richer.  We present a simple scenario where a symmetry renders the high scale of inflation natural, and hence leads to a detectable B mode polarization from the tensor vacuum fluctuations.  We note that this \emph{same} symmetry leads also to a series of interesting and, more importantly, \emph{correlated} observables.  In particular, we show how detectable resonant and equilateral non-Gaussianities may naturally arise, in concert with signals at gravitational wave interferometers (such as Advanced LIGO/VIRGO).

At the heart of our study is a rather simple and general observation: the inflaton field should couple not only to itself, but also to ``matter'' (non-inflationary) sectors.\footnote{In some cases, ``matter'' couplings can lead to dangerous loop corrections which spoil the flatness of the inflaton potential, if they are too strong.  However, this need not be the case in the most robust and appealing inflation models -- they rely on some symmetry which protects the slow roll parameters and strongly constrains the allowed interactions.}  Indeed, the presence of such couplings would seem to be a pre-requisite for successful reheating.  Although ``matter'' couplings are apparently a ubiquitous feature of realistic models, their relevance for the phenomenology of the primordial cosmological scalar/tensor fluctuations has only recently been appreciated.

In this work, we explore the consequences of ``matter'' couplings in a particularly natural model: inflation driven by a pseudo-scalar axion.  In this case, the inflaton enjoys a continuous shift symmetry which, although slightly broken, nevertheless protects the flatness of the potential from receiving unacceptably large loop corrections; see \cite{natural,natural1.5,Freese:1994fp,Kinney:1995cc,natural2,extranatural,2-flation,N-flation,N-flation2,monodromy,monodromy2,kaloper,mixing2,lorenzo,dante,kallosh,kallosh2,misra,grimm} for model building.  This construction is especially appealing in the case of high scale inflation -- which leads to observable tensor modes -- since an infinite number of corrections need to be suppressed due to the fact that the inflaton travels over a super-Planckian distance \cite{lyth}.  In models of axion inflation, there is generically present a coupling to gauge fields of the form
\begin{equation}
\label{Lint}
 \mathcal{L}_{\mathrm{int}} = -\frac{\alpha}{4 f} \varphi F_{\mu\nu}\tilde{F}^{\mu\nu} \, ,
\end{equation}
where $\varphi$ is the inflaton, $F_{\mu\nu} = \partial_\mu A_\nu - \partial_\nu A_\mu$ is the field strength associated with some $U(1)$ gauge field $A_\mu$, $\tilde{F}^{\mu\nu} = \eta^{\mu\nu\alpha\beta}F_{\alpha\beta}/(2\sqrt{-g})$ is the dual field strength (the generalization to non-Abelian groups is straightforward), $f$ is the axion decay constant and $\alpha$ is a dimensionless parameter.  Notice that, upon integration by parts this is manifestly a derivative coupling and therefore does not induce perturbative corrections to the inflaton potential.  The coupling (\ref{Lint}) has important implications for both gravitational waves and also non-Gaussianity.

It is natural to split up our discussion of the phenomenology of the coupling (\ref{Lint}) according to the relevant observational scales.  First, consider CMB/LSS scales.  There are two important sources of non-Gaussianity in axion inflation:
\begin{enumerate}
 \item Couplings of the type (\ref{Lint}) to non-Abelian gauge groups can induce, via instanton effects, non-perturbative oscillatory contributions to the axion potential.\footnote{The origin of these oscillations is somewhat different in string theory constructions, coming from ED1 corrections to the K\"ahler potential, but the end effect is equivalent.} These in turn lead to very distinct oscillations in the scalar power spectrum and also non-Gaussianity of the resonant type \cite{monodromy2,pajer,leblond,monodromyNG,chen2}.
 \item The homogeneous dynamics of the inflaton can lead to nonperturbative production of classical gauge field fluctuations due to the coupling (\ref{Lint}).  These produced gauge field fluctuations, in turn, source scalar inflaton perturbations by \emph{inverse decay}: $\delta A + \delta A \rightarrow \delta\varphi$.  This effect leads to a large equilateral contribution to the bispectrum, as was first realized in \cite{ai}.
\end{enumerate}
In section \ref{sec:CMB}, we review the computation of both resonance and inverse decay effects using the in-in formalism, showing how both effects can coexist in the spectrum and bispectrum.  We also generalize previous studies \cite{ai,ai2} to allow for the possibility of interactions which explicitly break the underlying shift symmetry.  (Such interactions are actually present in certain string theory constructions; more on this later.)

Next, let us consider the phenomenology on much smaller scales, generated around $10$ to $20$ e-foldings before the end of inflation. Since the produced gauge field fluctuations contribute to the anisotropic stress tensor, they provide an important new source of gravitational waves which is much larger than the usual one generated by vacuum fluctuations. Recently Cook and Sorbo \cite{GWlorenzo} have realized that this signal should be visible at interferometers such as Advanced LIGO/VIRGO and Einstein Telescope.\footnote{See also \cite{parity} for a discussion of the effects of this additional source of gravitational waves on CMB scales.  See \cite{GWeva} for a discussion of gravitational waves from particle/string production effects in different models.} We continue their analysis by taking into account the effect of the strong backreaction regime that necessarily  characterizes the late stage of inflation. We show that in the regime relevant for observations at interferometers, a copious production of gauge fields leads to about $10$ additional e-foldings of inflation as compared to the case where the gauge field coupling is absent.
As a consequence, the gravity wave modes left the horizon when the inflaton field was in a less flat part of the potential (at least, for the simplest monomial potentials for which this effect has been studied) and hence a larger population of gauge modes was present. This
enhances  the gravitational waves signal, and makes it  detectable  for the \emph{same} range of parameters that lead to observable non-Gaussianity, when one assumes the most natural values for the duration of inflation. Specifically, if one requires that the large  CMB scales left the horizon about $60$ e-foldings before the end of inflation - which, as we show in subsection \ref{subsec:reheating}, is the most natural expectation given the coupling (\ref{Lint}) - then backreaction effects enhance the gravitational wave signal to a level that can be observed at Advanced LIGO/VIRGO.  It is  worth pointing out that tensor modes produced by inverse decay encode information about the interactions of the inflaton unlike those produced by vacuum fluctuations, which depend only on the energy of the background.

It is interesting to note that the \emph{same} coupling (\ref{Lint}) which leads to interesting non-Gaussianity and gravitational wave signals will also provide a natural decay channel for the inflaton.  We provide a cursory discussion of the physics of (p)reheating in axion inflation, arguing that the decay of the inflaton is extremely efficient in the observationally interesting regime.

Throughout the first part of this paper, we work in a very general effective field theory context.  However, because inflationary model-building is sensitive to Ultra-Violet (UV) physics, it makes sense to explore the embedding of QFT models of inflation into a more complete framework, such as string theory. In particular we compute the size of the coupling \eqref{Lint} in a class of axion monodromy inflation in IIB string theory \cite{monodromy,monodromy2,dante}. A model independent coupling is present to the gauge fields living on the world volume of the NS5-brane that generates the monodromy. Whether this is large enough to trigger the inverse decay phenomenology depends on the size of $g_{s}C_{0}$ (the string coupling times the RR zero form). For $g_{s}C_{0}\lesssim\mathcal{O}(1)$ as expected in a generic perturbative flux compactification, the coupling \eqref{Lint} is too small. On the other hand, although it is less common (we found it in $.5\%$ of the toy models we considered), nothing forbids $g_{s}C_{0}\gtrsim7$ in which case signatures at interferometers and in the CMB would be present. For completeness we computed the model dependent coupling to the gauge fields on a D5-branes wrapping the cycle defining the axion. This is naturally of the order of the current experimental bound on inverse decay non-Gaussianity, and its precise value depends on volume moduli.

Although we will work in the context of a specific class of models, our key observation -- the importance of inflaton-matter couplings for non-Gaussian phenomenology -- is of a much broader nature.  In \cite{sarah} it was argued that the scenario of \cite{ai,ai2} is couched within a more general class of theories which exhibit the so-called ``feeder'' mechanism.  (Another example was explored in detail in \cite{pp1,pp2,pp3,NGreview}.)  Indeed, the example interaction (\ref{Lint}) serves to highlight an important point: even very simple QFT models of inflation can lead to surprisingly complex dynamics and rich phenomenology.

The paper is organized as follows. In section \ref{sec:bkg} we study the homogeneous background dynamics of the model (\ref{Lgen}). In section \ref{sec:CMB} and \ref{sec:LIGO} we study the scalar cosmological perturbations on scales relevant for CMB/LSS and interferometers, respectively. Finally, after reviewing in section \ref{s:review} the string theory construction of axion monodromy inflation and the challenges it faces, in section \ref{s:modin} we identify cases where the coupling to gauge fields is large enough to trigger interesting phenomenological effects.


\subsection{Effective Field Theory of Axion Inflaton}\label{s:EFT}

To be concrete, let us assume that the effective field theory of the inflaton is characterized by a (mildly broken) shift symmetry
\begin{equation}
\label{shift_symm}
 \varphi \rightarrow \varphi + \mathrm{const} \, .
\end{equation}
Such a symmetry could arise because $\varphi$ is a Pseudo-Nambu-Goldstone-Boson (PBNG) associated with a global symmetry that is spontaneously broken at some scale $f$.  In the context of string theory, the axion arises instead from the dimensional reduction of a two-form over an internal two-cycle.  In this case, the symmetry (\ref{shift_symm}) has its origin in the gauge symmetry of the two-form.  Models with an underlying shifty symmetry have attracted a considerable amount of interest, partially motivated by the fact that such theories can naturally lead to an observable tensor-to-scalar ratio \cite{natural,natural1.5,natural2,extranatural,2-flation,N-flation,N-flation2,monodromy,monodromy2,kaloper,mixing2,lorenzo,dante,kallosh,kallosh2,misra,grimm}.

The symmetry (\ref{shift_symm}), although valid to all orders in perturbation theory, is generically broken by non-perturbative effects to a discrete subgroup $\varphi \rightarrow \varphi + 2 \pi f$, leading to a periodic contribution to the effective potential of the form
\begin{equation}
\label{Vnp}
  V_{\mathrm{np}}(\varphi) = \Lambda^4 \cos\left(\frac{\varphi}{f}\right) + \dots
\end{equation}
where $f$ is the axion decay constant, $\Lambda$ is a non-perturbatively generated scale and $\dots$ denotes (sub-dominant) higher harmonics.  The original natural inflation model \cite{natural} exploited the potential (\ref{Vnp}) to drive inflation.  However, this simple scenario is compatible with observation only when $f\gsim 4 M_p$ \cite{natural2}, a regime that seems problematic:  In the case of a PBNG, $f>M_p$ would suggest a global symmetry that is broken \emph{above} the quantum gravity scale, where conventional QFT is presumably not valid \cite{Kallosh:1995hi,big_f,extranatural};  moreover, values $f>M_p$ do not seem possible in a controlled limit of string theory \cite{big_f}.

In addition to (\ref{Vnp}), one may also incorporate additional ingredients which break the symmetry (\ref{shift_symm})
explicitly at tree level and lead to a \emph{monodromy}: the would-be periodic direction $\varphi$ is ``unwrapped'' \cite{monodromy,monodromy2,kaloper,mixing2}.  In many interesting scenarios, explicit symmetry breaking ingredients induce a power-law potential of the form
\begin{equation}
\label{Vsr}
  V_{\mathrm{sr}}(\varphi) = \mu^{4-p} \varphi^p \, .
\end{equation}
The case $p=1$ is typical of string theory models \cite{monodromy,monodromy2} while $p=2$ arises from axion/4-form mixing \cite{kaloper,mixing2}.  

Notice that, generically, \emph{both} contributions (\ref{Vnp}) and (\ref{Vsr}) may be present:
\begin{equation}
\label{Vintro}
  V(\varphi) = V_{\mathrm{np}}(\varphi) + V_{\mathrm{sr}}(\varphi) \, .
\end{equation}
As the suffix sr (slow roll) indicates, it is often assumed that $V_{\mathrm{sr}}$ dominates the potential; one then obtains a realization 
of large-field inflation, even for $f \ll M_p$.  The potential (\ref{Vintro}) must be constant in the limit that the symmetry (\ref{shift_symm})
is exact.  Hence, we expect that the smallness of symmetry breaking effects will protect the slow roll parameters
\begin{equation}
\label{SR}
 \epsilon_V = \frac{M_p^2}{2}\left(\frac{V'}{V}\right)^2, \hspace{5mm} \eta_V \equiv M_p^2 \frac{V''}{V} \, ,
\end{equation}
from receiving unacceptably large radiative corrections.

The \emph{same} symmetry which protects the slow roll parameters also constrains the interactions of $\varphi$ and therefore has implications for 
non-Gaussianity.  In our analysis, a key role will be played by the presence of pseudo-scalar interactions (\ref{Lint}) between the inflaton and the gauge fields.  The interaction (\ref{Lint}) is consistent with the underlying symmetries and, from an effective field theory perspective, it \emph{must} be included.  The strength of the pseudo-scalar interaction (\ref{Lint}) is controlled by the axion decay constant $f$; absent fine-tuning we do not expect $\alpha \ll 1$.  In passing, notice that a coupling of the form (\ref{Lint}) to the visible sector provides a natural decay channel for the inflaton and hence may be considered phenomenologically desirable. In principle, one can also consider the coupling of the pseudo-scalar inflaton to fermions. However, this results in a helicity-suppressed decay rate, which vanishes in the limit of vanishing fermion mass. We expect that, if the inflaton decays into significantly lighter  fields, the decay into fermions can be neglected in comparison to that into gauge fields.

In the limit that the symmetry (\ref{shift_symm}) is exact, only the coupling (\ref{Lint}) to gauge fields is permitted (we note that the quantity $F\tilde{F}$ is a total derivative so that $\varphi\rightarrow\varphi+c$ shifts the Lagrangian by a surface term which has no impact on the classical equations of motion).
However, the explicit symmetry breaking ingredients which lead to the potential (\ref{Vsr}) may, in principle, lead to additional couplings.  A generic interaction Lagrangian which is compatible with gauge symmetry is
\begin{equation}
\label{Lintgen}
  \mathcal{L} \supset - \frac{B(\varphi)}{4}F^2 - \frac{C(\varphi)}{4} F\tilde{F}  \, .
\end{equation}
The smallness of the symmetry breaking -- which is required to ensure slow roll -- implies that $B(\varphi)$ must be nearly constant.  To quantify the slow variation of the function $B(\varphi)$, it is useful to define the following parameters
\begin{equation}
\label{SRB}
  \epsilon_B \equiv \frac{M_p^2}{2}\left(\frac{B'}{B}\right)^2, \hspace{5mm} \eta_B \equiv M_p^2 \frac{B''}{B} \, .
\end{equation}
where the prime denotes derivative with respect to $\varphi$.  The same logic implies that $C(\varphi)$ must be close to linear, and hence $C'(\varphi)$ should be nearly constant.  We therefore introduce an additional set of slow-variation parameters
\begin{equation}
\label{SRC}
 \epsilon_C \equiv \frac{M_p^2}{2}\left(\frac{C''}{C'}\right)^2, \hspace{5mm} \eta_C \equiv M_p^2 \frac{C'''}{C'} \, .
\end{equation}
(Again here the prime denotes derivative with respect to $\varphi$.)  Throughout this work we assume that $\epsilon_i,|\eta_i| \ll 1$ for $i = V, B,C$.

Putting everything together, and including mild symmetry breaking effects, we are led to consider the following effective field theory description of an axion inflaton
\begin{equation}
\label{Lgen}
  \mathcal{L} = -\frac{1}{2}(\partial \varphi)^2 - \mu^{4-p} \varphi^p - \Lambda^4 \cos\left(\frac{\varphi}{f}\right) 
  - \frac{B(\varphi)}{4}F^2 - \frac{C(\varphi)}{4} F\tilde{F} \, .
\end{equation}
The theory (\ref{Lgen}) subsumes nearly all known example of axion inflation and generalizes the model of \cite{ai,ai2} to incorporate also explicit symmetry breaking interactions, parametrized by the coupling functions $B(\varphi)$, $C(\varphi)$.

Throughout this paper, we assume a flat Friedmann-Lema\^itre-Robertson-Walker (FLRW) background geometry: $ds^2 = -dt^2 + a^2(t) d{\bf x}^2$ with scale factor $a(t)$.  We sometimes use conformal time $\tau = \int \frac{dt}{a}$.  Derivatives with respect to cosmic time are denoted as $\dot{f} \equiv \partial_t f$ and with respect to conformal time as $f' \equiv\partial_\tau f$.


\section{Background Evolution and Backreaction Effects}
\label{sec:bkg}

In this Section we study the homogeneous background evolution of the model (\ref{Lgen}).  We are interested in an inflaton potential of the form
\begin{equation}
\label{V}
  V(\varphi) = \Vsr(\varphi) + \Lambda^4\cos\left(\frac{\varphi}{f}\right) \, .
\end{equation}
We require that $\Vsr$ is sufficiently flat to drive slow roll inflation, while the oscillatory term may be treated as a small modulation.  In this Section, we disregard the oscillatory term in (\ref{V}), which has a small impact on the number of e-foldings of inflation.\footnote{The oscillatory term in (\ref{V}) is important, however, for the cosmological fluctuations; see Section \ref{sec:CMB}.}   Instead, we focus on the effect of the gauge field couplings in (\ref{Lgen}).  We will see shortly that the production of gauge fluctuations can lead to important backreaction effects on $\phi(t) = \langle\varphi(t,{\bf x}) \rangle$.  These effects are the subject of this Section. The discussion is divided in three Subsections.  In Subsection \ref{subsec:gaugeprodn} we review the production of the gauge quanta due to  the motion of the inflaton, generalizing previous results.  Some technical details of the derivation are relegated to Appendix A.  In Subsection \ref{subsec:classicalevolution} we show  that, if the inflaton-gauge field coupling is sufficiently large to  lead to observable non-Gaussianity from inverse decay, then the produced quanta affect the inflaton dynamics during the last $N \sim 20$ e-foldings of inflation.  This effect produces a change of the predicted scalar spectral index $n_s$ and tensor-to-scalar ratio $r$ which could be effectively ascribed to a shift of $\Delta N \sim 10$ of the number of e-foldings from horizon crossing for CMB scales to the end of inflation.  In Subsection \ref{subsec:reheating} we briefly discuss the endpoint of inflation, commenting in particular on (p)reheating.


\subsection{Gauge Field Production}
\label{subsec:gaugeprodn}

As first  shown in \cite{lorenzo}, the homogeneous dynamics of the inflaton leads to an important instability for the gauge field, $A_\mu$.  To see this effect, we consider the equation of motion for gauge field fluctuations in the background of the homogeneous inflaton $\phi(t) \equiv \langle\varphi(t,{\bf x})\rangle$.  We introduce a canonical field variable by the rescaling\footnote{In principle this introduces a kinetic mixing with the inflaton, but since we are assuming $\epsilon_{B}\ll1$ this is a subleading effect which we therefore neglect.}
\begin{equation}
  \tilde{A}_i(t,{\bf x}) \equiv \sqrt{B(\phi(t))} \, A_i(t,{\bf x})\,
\end{equation}
and decompose $\tilde{A}_i(t,{\bf x})$ as
\begin{equation}
\label{Adecomp}
  \tilde{A}_i(\tau,{\bf x}) = \sum_{\lambda=\pm} \int \frac{d^3k}{(2\pi)^{3/2}} 
  \left[ \epsilon_{i}^{\lambda}({\bf k}) a_\lambda({\bf k}) \tilde{A}_\lambda(\tau,{\bf k}) e^{i{\bf k}\cdot {\bf x}} +\mathrm{h.c.} \right] \, ,
\end{equation}
where $\vec{\epsilon}_\lambda$ are circular polarization tensors (see \cite{ai2} for more details) and the annihilation/creation operators
of the gauge field obey
\begin{equation}
  \left[a_{\lambda}({\bf k}), a_{\lambda}^\dagger(\bf k')\right] = \delta_{\lambda\lambda'}\delta^{(3)}\left({\bf k}-{\bf k'}\right) \, .
\end{equation}

The c-number mode functions obey the follow equation of motion
\begin{equation}
\label{Aeqn}
 \left[ \frac{\partial^2}{\partial\tau^2} + k^2 + M^2_A(\tau) \pm \frac{2 k \xi}{\tau}  \right] \tilde{A}_{\pm}(k,\tau) = 0 \, ,
\end{equation}
where 
\begin{eqnarray}
  \xi &\equiv& \sqrt{\frac{\epsilon_\phi}{2}} M_p \frac{C'}{B} \mathrm{sign}\left(\dot{\phi}\right)  \, , \\
  M^2_A(\tau) &\cong& -\frac{\sqrt{\epsilon_\phi \epsilon_B}}{\tau^2} \mathrm{sign}\left(B'\dot{\phi}\right) \, .
\end{eqnarray}
and where prime denotes derivative with respect to $\phi$ and we have introduced the slow roll parameter
\begin{equation}
 \epsilon_\phi \equiv \frac{\dot{\phi}^2}{2 H^2 M_p^2} \, .
\end{equation}
During the conventional inflationary regime we have the usual relations $\epsilon_\phi \cong \epsilon_V \cong -\frac{\dot{H}}{H^2}$ (see Eqn.~\ref{SR}).  However, when backreaction effects are important these different definitions need not coincide.

The quantity $M^2_A(\tau)$, which was not present in \cite{lorenzo,ai,ai2}, arises due to the explicit shift symmetry breaking interactions (\ref{Lintgen}); this term vanishes when the shift symmetry (\ref{shift_symm}) is exact and $B(\varphi) = \mathrm{const}$.  Evidently, $M^2_A(\tau)$ is proportional to slow-variation parameters. As we will see shortly, this contribution to the effective frequency of the gauge field fluctuations does not have any significant impact on their dynamics.  On the other hand, the term in (\ref{Aeqn}) proportional to $\xi$ plays a very important role.  

Without loss of generality, we can take $\xi > 0$.  Moreover, we will be most interested in $\xi\gsim 1$.  In this case the $\lambda=+$ polarization state experiences a tachyonic instability for $k / (a H) \lsim 2\xi$, leading to an exponential growth of fluctuations.  (The $\lambda=-$ polarization instead remains in its vacuum and can be disregarded.)  More specifically, for $\xi = \mathcal{O}\left( 1 \right)$, the gauge field modes  are in the vacuum state during most of the sub-horizon regime, and the tachyonic instability becomes appreciable only close to horizon crossing. The growth ceases when the mode becomes much greater than the horizon, and then the energy density in that mode decreases due to the expansion of the universe. Therefore, at any given moment during inflation, only gauge modes of comparable size to the horizon are relevant. 

Typically, the parameter $\xi \propto \dot{\phi} / H$ increases during inflation. However, the time variation of $\xi$ appears only at higher order in  $\epsilon$, $\eta$. We can therefore treat $\xi$ as an adiabatically evolving parameter: we treat it as a constant during the interesting part of the evolution of each mode (since $\vert \Delta \xi \vert \ll \xi$ over $\Delta t = {\rm few} \times H^{-1}$). However, when comparing two well separated stages of inflation, we take into account that $\xi$ can be different between them, leading to a different amount of production. We note that $\xi$ instead varies significantly after inflation, so that we do not expect that treating $\xi$ as a constant provides a good approximation to the gauge field production during this stage.

In Appendix A we derive an analytic expression for the properly-normalized solutions of (\ref{Aeqn}), in the approximation that $\epsilon_\phi$, $\epsilon_B$ and $\xi$ may be treated as constant (slow roll approximation).  We show that the solutions admit the simple representation:
\begin{equation}
\label{Amode}
 \tilde{A}_+(k,\tau) \cong \frac{1}{\sqrt{2k}}\left(\frac{k}{2\xi a H}\right)^{1/4} e^{\pi \xi - 2\sqrt{2\xi k / (aH)}} \, .
\end{equation}
 We note the exponential enhancement, $e^{\pi \xi}$, in (\ref{Amode}), which reflects the underlying tachyonic instability.  Equation (\ref{Amode}) coincides (up to a change in normalization) with the result that was first derived in \cite{lorenzo} for the case $B=1$, $C\propto\varphi$. As first discussed  in \cite{ai}, this approximation is valid in  the region $(8\xi)^{-1} \lsim -k\tau \lsim 2\xi$ that accounts for most of the power in the produced fluctuations (we stress that this phase space is nontrivial for $\xi \gsim \mathcal{O}(1)$, which we assume throughout; the production of gauge fluctuations is uninterestingly small for $\xi < 1$).  The result (\ref{Amode}) highlights our previous claim that the explicit symmetry breaking interactions (\ref{Lintgen}) do not significantly impact the production of gauge fluctuations in axion inflation.  See Appendix A for more details.

The modes (\ref{Amode}) are real-valued (up to an irrelevant constant phase); this  leads to an important simplification when we compute non-Gaussian correlators.  The fact that the c-number modes are real-valued implies that the produced gauge field fluctuations are commuting variables, to a very good approximation.  That is, we have
\begin{equation}
\label{Acommute}
 \left[ \partial_\tau A_i(\tau,{\bf x}), A_j(\tau,{\bf x'})  \right] \approx 0 \, ,
\end{equation}
where it is understood that only the produced fluctuations  with $-k\tau < 2\xi$ are relevant (the modes with $-k\tau \gg 2\xi$ remain in their vacuum and their effect is renormalized away).  Equation (\ref{Acommute}) reflects the essentially classical nature of the tachyonic production of gauge fluctuations (we note the analogy with the decoherence and  semiclassicality of the standard cosmological perturbations due to the expansion of the universe \cite{Polarski:1995jg}).


\subsection{Classical Evolution}
\label{subsec:classicalevolution}

In this subsection, we study the backreaction of the produced gauge field fluctuations (\ref{Amode}) on the homogeneous dynamics of the inflaton $\phi(t)$.  The smallness of the parameters (\ref{SRB}) and (\ref{SRC}) implies that $B(\varphi)$ should be very close to a constant (that can be absorbed into the normalization of the gauge field) and $C(\varphi)$ should be very close to linear.  Hence, for simplicity, we focus on the case 
\begin{equation}
 B(\varphi) \equiv 1, \hspace{5mm} C(\varphi) \equiv \frac{\alpha}{f} \varphi \, ,
\;\; \Rightarrow \;\; \xi = \frac{\alpha \, \dot{\phi}}{2 f H}
\label{B-C-xi-simple}
\end{equation}
for the remainder of this section. We also focus the discussion to the  string theory axion monodromy inflationary potential (see Section \ref{s:modin}) 
\begin{equation}
\label{Vstring}
 V_{\mathrm{sr}}(\varphi) = \mu^3  \left[ \sqrt{\varphi^2 + \varphi_c^2} - \varphi_c \right]
\end{equation}
During inflation we have $\varphi \gg \varphi_c$ so that $V(\varphi) \approx \mu^3|\varphi|$.  (The parameter $\varphi_c$ is, however, relevant during reheating.)

We introduce the physical ``electric'' and ``magnetic'' fields as
(despite the terminology, the gauge field $A_\mu$ need not correspond to the Standard Model photon)
\begin{equation}
 \vec{B} = \frac{1}{a^2} \vec{\nabla}\times \vec{A}, \hspace{5mm} \vec{E} = -\frac{1}{a^2} \vec{A}' \, .
\end{equation}

The gauge field fluctuations (\ref{Amode}) are produced at the expense of the kinetic energy of the homogeneous inflaton $\phi(t)$, introducing a new source of dissipation into the  equation of motion of $\phi$.  Moreover, these produced fluctuations contribute to the energy density of the universe and thus modify Friedmann equation.  These effects are encoded in the mean field relations
\begin{eqnarray}
&&\ddot{\phi} + 3 H \dot{\phi} + V' = \frac{\alpha}{f} \langle \vec{E} \cdot \vec{B} \rangle \, , \nonumber\\
&& 3 H^2 = \frac{1}{M_p^2} \left[ \frac{1}{2} \dot{\phi}^2 + V + \frac{1}{2} \langle \vec{E}^2 + \vec{B}^2 \rangle \right] \, , 
\label{eom-back}
\end{eqnarray}
and, using (\ref{Amode}), one finds  \cite{lorenzo}
\begin{equation}
\langle \vec{E} \cdot \vec{B} \rangle  \simeq - 2.4 \cdot 10^{-4}   \, \frac{H^4}{\xi^4} \, {\rm e}^{2 \pi \xi} \;\;,\;\; \langle \frac{\vec{E}^2+\vec{B}^2}{2} \rangle \simeq 1.4 \cdot 10^{-4} \frac{H^4}{\xi^3} {\rm e}^{2 \pi \xi} \, . \label{backeqns}
\end{equation}
Therefore  \cite{ai,ai2} 
\begin{eqnarray}
 \xi^{-3/2} \, {\rm e}^{\pi \xi} \ll  79 \, \frac{\dot{\phi}}{H^2}
   &\Rightarrow& {\rm negligible \; backreaction \; on \; } \phi \; {\rm  eq.} \nonumber\\
 \xi^{-3/2} \, {\rm e}^{\pi \xi} \ll  146 \, \frac{M_p}{H}
  &\Rightarrow&  {\rm negligible \; backreaction \; on \; Friedmann \; eq.} 
\label{noback}
\end{eqnarray} 
During the standard slow roll regime $\dot{\phi} \simeq \sqrt{2 \epsilon_\phi} \, H \, M_p \ll H M_p$, and the first condition is the more stringent than the second. If we take the standard result for the spectrum of primordial perturbations, $P_\zeta^{1/2}  = H^2 / \left( 2 \pi \dot{\phi} \right) \simeq 5 \cdot 10^{-5}$, this condition reads $\xi <  4.7$. 

In \cite{ai,ai2} it was shown that $\xi \simeq 2.5$ leads to the standard result for the power spectrum (up to a subdominant correction due to the modes generated by inverse decay), and to observable (but not yet ruled out) non-Gaussianity from the inverse decay of the gauge quanta into inflaton quanta. We see that backreaction is indeed negligible for such value. However,  we note that $\xi$ typically  increases during inflation. For concreteness consider a linear inflaton potential (such is the case for monodromy), then $\xi \propto 1 / \phi$ during the standard slow roll regime. Let us assume that  $\xi = 2.5$ at $\vert \phi \vert  \simeq 11 M_p$, which is the value of the inflaton about $60$ e-foldings before the end of inflation, under the standard slow roll assumption. We then find that $\xi = 4.7$ is reached at $\vert \phi \vert  \simeq 5.9 \, M_p$. This happens before inflation ends. As discussed in \cite{lorenzo}, the production of gauge quanta results in an additional friction on the inflaton motion that, in the current context, prolongs the duration of inflation. 

We evolved numerically the two equations~(\ref{eom-back}), keeping into account the backreaction term in the equation for the scalar field. We still disregarded the energy density of the gauge quanta in the Friedmann equation; in this way this equation remains a simple algebraic equation for $H$; the right panel of Figure \ref{Fig:back2} shows that this is a good approximation all the way to the end of inflation. 
We choose parameters such that, initially, we are in the standard slow roll regime adopted in  \cite{ai,ai2}. Therefore, the result for the power spectrum, and for $f_{\rm NL}$ obtained in \cite{ai,ai2}, apply (see Subsection \ref{subsec:n-pt}). We note that the coupling between the inflaton and the gauge field can be given in terms of the initial values of $\xi$ and $\phi$:
\begin{equation}
 \xi \Big\vert_{\phi_{CMB}} = 2.5 \;\;\Rightarrow\;\;
\frac{\alpha}{f} = 5 \,  \frac{\vert \phi_{CMB} \vert}{M_p^2} \, .
\label{mu-alphaf}
\end{equation}
Here and in the following, the suffix ``CMB'' denotes the value of quantities when the large scale CMB perturbations left the horizon (when this statement needs to be quantified, we refer to the WMAP  pivot scale \cite{wmap7} $k=0.002 \, {\rm Mpc}^{-1}$).

We obtained $\vert \phi_{CMB} \vert  \simeq 9.9 \, M_p \,$; this is smaller than the  value ($\vert \phi_{CMB} \vert  \simeq 10.9\, M_p$) leading to $60$ e-foldings of inflation without gauge production, confirming  that the backreaction of the produced quanta increases the amount of inflation.

\begin{figure}[h]
\begin{center}\includegraphics[width=0.5\textwidth]{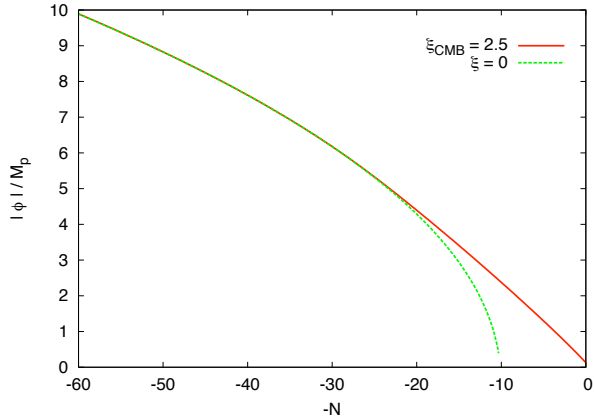}\end{center}
\caption{Evolution of the inflaton as a function of the number of e-foldings to the end of  inflation, starting from $\vert \phi_{CMB} \vert  = - 9.9 \, M_p $,  with (red solid line) and without (green dashed line) the coupling to gauge fields. For the first line, the strength of the inflaton-gauge field coupling is chosen so to lead to observable non-Gaussianity from inverse decay. For the second line, we have shifted the number of e-foldings to make manifest that  the two evolutions coincide at early times.
}
\label{Fig:back}
\end{figure}

In figure \ref{Fig:back} we show the evolution of the inflaton field as a function of the number of e-foldings to the end of inflation for $\xi_{CMB} = 2.5$ (red solid curve) and for $\xi=0$ (green dashed line), i.e~the standard slow-roll case. The backreaction of the produced quanta on the background evolution becomes noticible during the last $\sim 25$ e-foldings of inflation, while it is negligible at earlier times. The two trajectories reach $\phi=0$ at different times, showing that the backreaction increases the duration of inflation by about $10$ e-foldings.

\begin{figure}[h]
\begin{center}
\includegraphics[width=0.45\textwidth]{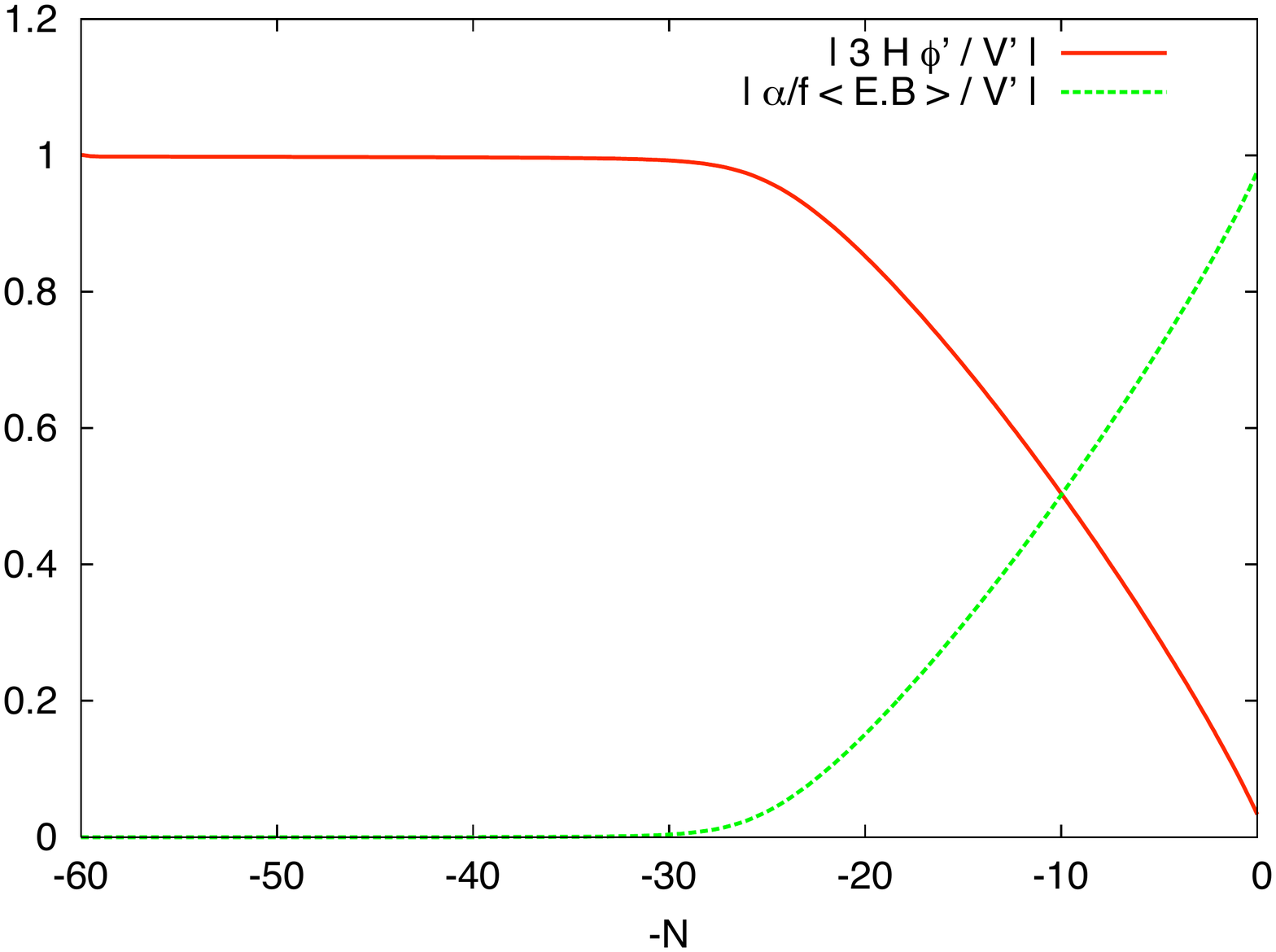}
\includegraphics[width=0.45\textwidth]{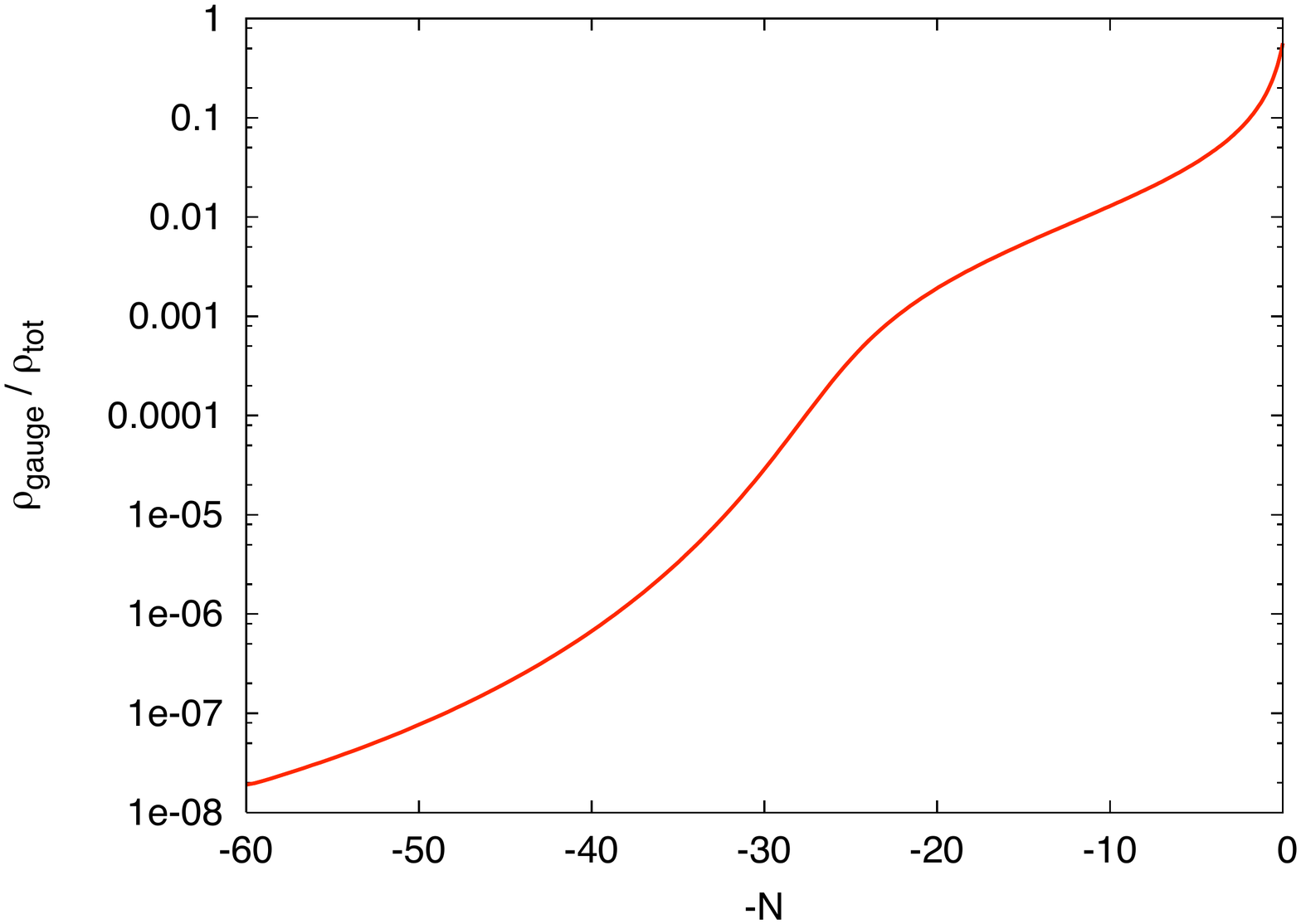}
\end{center}
\caption{Left panel: Friction terms in the equation of motion for $\phi$. Right panel: relative strength of the energy density of the produced quanta; this term is neglected in the numerical evolution of the background equations.
}
\label{Fig:back2}
\end{figure}

This change in behavior during the last $\sim 25$ e-foldings of inflation is also visible in the left panel of figure \ref{Fig:back2}, where we show the evolution of the two friction terms in the inflaton equation as a function of the number of e-foldings to the end of inflation. The standard Hubble friction controls the earliest stages, but the system gradually evolves towards a regime in which the backreaction of the produced gauge quanta dominates the evolution. Namely, the system approaches the strong backreaction regime studied in \cite{lorenzo}. Let us stress that in the our case the observable cosmological fluctuations were produced during the phase of standard slow roll inflation (and therefore, the results of  \cite{ai,ai2} apply), while the strong backreaction regime of  \cite{lorenzo} here is reached only in the last $\mathcal{O}(10)$ e-foldings of inflation. In the right panel of figure \ref{Fig:back2} we show the relative contribution to the energy density of the gauge quanta to the total energy density. This term was neglected in the numerical evolution of the system (\ref{eom-back}). The Figure confirms that this is a valid assumption.


\subsection{Reheating}
\label{subsec:reheating}

We conclude this Section with a brief comment on reheating in our model.  It is interesting that in this model the \emph{same} coupling $\varphi F\tilde{F}$ which lead to non-Gaussianity will also provide a natural decay channel for the inflaton.  We expect reheating into the gauge bosons $A_\mu$ to be very efficient whenever particle production effects are important.  This is illustrated in the right panel of Fig.~\ref{Fig:back2}.  There we see that, already by the end of inflation, the energy density in gauge field fluctuations is comparable to that of the inflaton.  At this point it may be expected that the condensate $\phi(t)$ is efficiently destroyed and thermal equilibrium may set in rather quickly. 

A detailed account of the dynamics of (p)reheating in the model (\ref{Lgen}) goes beyond the scope of this paper.  Here we briefly discuss the perturbative decay of an axion inflaton, anticipating that such an analysis might be viewed as a lower bound on the efficiency of reheating.  The perturbative decay rate into gauge fields associated with the coupling (\ref{Lint}) is
\begin{equation}
 \Gamma_{\varphi\rightarrow AA} = \frac{\alpha^2 m_\phi^3}{64 \pi f^2} \, .
\label{pert-dec}
\end{equation}
The inflaton mass is obtained from the potential $V_{\rm sr}$.  For string theory monodromy inflation models this is given by (\ref{Vstring}) where, as we discuss in Section \ref{s:modin}, $\varphi_c = {\rm O } \left( 10^{-1} \right) M_p $ (we note that, indeed, $V_{\rm sr}  \simeq \mu^3 |\phi| $ during inflation). This leads to $m_\phi \sim 5 \cdot 10^{-5} \, M_p$. Taking   $\alpha/f$ as in (\ref{mu-alphaf}), and equating $ \Gamma_{\varphi\rightarrow AA} = H $, we obtain the estimate $T_{\rm rh} = {\rm O } \left( 10^{12} \right) {\rm GeV} $ on the reheating temperature of the thermal bath generated by the perturbative decay.  As mentioned above, for the values of the coupling we are interested in, the energy density of the gauge quanta becomes comparable to that of the inflaton already at the end of inflation.  We expect that this will lead to additional backreaction effects beyond the mean field backreaction  of  $\langle E \cdot B \rangle $ and $\langle E^2 + B^2 \rangle$, which is the only effect that enters in (\ref{eom-back}). Specifically, we expect strong rescattering effects that should lead to a much quicker thermalization than the perturbative rate suggests. As for preheating of scalars, we believe that it would be worthwhile to study these processes through lattice simulations.

It is worth mentioning that we have implicitly assumed that the inflaton-gauge interaction in (\ref{Lint}) is the strongest interaction of the inflaton to any gauge field. Pseudoscalars have also a natural coupling to fermions $\Delta {\cal L} = - \frac{C \, m_\psi}{f} \, \phi  \, {\bar \psi} \gamma_5 \psi$, where $C$ is a model dependent quantity which, in the spirit of an effective field theory, should not be expected to be $\ll 1$. The associated decay into fermions is helicity suppressed, $\Gamma_{\phi \rightarrow \psi \psi} \simeq \left( C / f \right) m_\phi \, m_\psi^2 / \left( 8 \pi \right)$ for $m_\psi \ll m_\phi$. This is parametrically smaller than the decay rate into gauge fields if $m_\psi \ll m_\phi$.

The detailed history of reheating is necessary to relate the size of any given mode to the number of e-foldings $N$ before the end of inflation at which this mode exited the horizon. This relation is given by \cite{Nk}:
\begin{equation}
N \left( k \right) = 62 - {\rm ln } \frac{k}{H_0 a_0} - {\rm ln } \frac{10^{16} \, {\rm GeV}}{V_{CMB}^{1/4}} + {\rm ln } \frac{V_{CMB}^{1/4}}{V_{\rm end}^{1/4}} - \frac{1}{3} {\rm ln } \frac{V_{\rm end}^{1/4}}{\rho_{\rm rh}^{1/4}} \, ,
\end{equation}
where the suffices CMB, ``end'', ``rh'', and $0$ denote  horizon crossing during inflation, the end of inflation, the conclusion of reheating, and the present epoch, respectively. For $\xi_{CMB} = 2.5$, and $\vert \phi_{CMB} \vert = 9.9 M_p$, this relation evaluates to
\begin{equation}
N \left( k \right) \simeq 61.7 - {\rm log } \frac{k}{0.002 \, {\rm Mpc}^{-1}} - \frac{1}{3} \, {\rm ln } \frac{10^{16} \, {\rm GeV}}{\rho_{\rm rh}^{1/4}} \, .
\label{e-folds}
\end{equation}
(Notice that, due to the logarithmic dependence, this relation will change only very slightly if $\xi_{CMB}$ and $\phi_{CMB}$ differ from the specific  values considered here.) We have normalized the wavenumber $k$ to the WMAP pivot scale \cite{wmap7}.

The dependence on $\rho_{\rm rh}$ in this relation is due to the assumption that, between the end of inflation and the completion of reheating, the massive  inflaton field is coherently oscillating about the minimum of its potential, resulting in a matter dominated stage. We see that, in general, a longer period of matter domination during inflation results in a lower value of $N$. Above, we have seen that the perturbative decay rate (\ref{pert-dec}) leads to $T_{\rm rh} \sim 10^{12} \, {\rm GeV}$. Inserting this in (\ref{e-folds}), we obtain $N \simeq 59$ for the CMB pivot scale. This however assumes matter domination of the inflaton oscillations before the perturbative decay takes place. We however see that, already by the end of inflation, the  gauge fields carry a non-negligible fraction of the total energy density. One may therefore expect that the equation of state of the universe will have an intermediate value between that of radiation and matter \cite{Podolsky:2005bw}. This would lead to a number of e-foldings between $59$ and $62$ for the CMB pivot scale.

\section{Fluctuations at CMB/LSS Scales}
\label{sec:CMB}

In this Section, we study the scalar cosmological fluctuations in the model (\ref{Lgen}) on CMB/LSS scales, corresponding roughly to scales which left the horizon $\sim 55-60$ e-foldings before the end of inflation.  In section \ref{sec:bkg}, we have seen that backreaction effects are negligible during this regime.  Even though backreaction effects are negligible, the production of gauge field fluctuations still has an important impact on the observable cosmological fluctuations, via inverse decay processes \cite{ai,ai2}.  Here, we re-visit the computation of \cite{ai,ai2} using the in-in formalism and allowing for the possibility of explicit symmetry breaking interactions.  We also account for the possibility of resonance effects, showing how these ``add up'' with the fluctuations from inverse decay.

This Section is divided into four Subsections.  In Subsection \ref{subsec:in-in} we derive the interaction Hamiltonian for the theory (\ref{Lgen}), accounting for all relevant effects.  Technical details are provided in Appendix B.  This effective action is the key input for a computation of the $n$-point correlation functions using the in-in formalism.  In Subsection \ref{subsec:n-pt} we briefly outline the computation of the spectrum and bispectrum using the in-in method.  Because the technical details are quite involved, we expand on this analysis in Appendices C and D.  Finally, in Subsection \ref{subsec:observables} we study observational consequences and delineate interesting regions of parameter space.

\subsection{Effective Action for the Perturbations and In-In Calculation}
\label{subsec:in-in}

We are interested in a potential of the form (\ref{V}), where the oscillatory term can be treated as a small correction and backreaction effects are negligible.  It is useful to introduce a parameter 
\begin{equation}
\label{b}
  b \equiv \frac{\Lambda^4}{f \, V'_{\mathrm{sr}}(\phi_\star)} \, ,
\end{equation}
where $\phi_\star$ is $\phi(t)$ evaluated at the moment $t_\star$ when some relevant pivot scale $k_\star$ left the horizon. If we assume $b<1$ so that \eqref{V} is monotonic, then the data require \cite{monodromy2} $b \ll 1$ so that a perturbative expansion in $b$ is justified. We will assume this regime throughout the paper.  The solution for the background evolution of the inflaton, $\phi(t) \equiv \langle \varphi(t,{\bf x})\rangle$, can be computed to leading order in $b$, with the result \cite{pajer}
\begin{equation}
\label{phi_bkg}
  \phi(t) = \phi_0(t) - \frac{3 b f^2}{\sqrt{2 \epsilon_\star}\, M_p} \sin\left(\frac{\phi_0(t)}{f}\right) \, ,
\end{equation}
where $\epsilon_\star \equiv -\dot{H}/H^2$ is a slow roll parameter,\footnote{In this Section we work in the regime where backreaction effects are negligible and it is not important to distinguish between $\epsilon_\phi$, $\epsilon_V$ and $-\dot{H}/H^2$.} evaluated at $t_\star$.  We have also introduced $\phi_0(t)$, which is the solution for the background scalar field in the absence of modulations, that is for $b=0$.  (Obviously, $\phi_0(t)$ depends on the specific choice of $\Vsr$.)  The slow roll parameters also admit an expansion in $b\ll 1$.  In particular, we can write $\epsilon = -\dot{H} / H^2$ as:
\begin{eqnarray}
 && \epsilon = \epsilon_0 + \epsilon_1 + \cdots \, ,\\
 && \epsilon_1 = -3 b \frac{f}{M_p} \sqrt{2\epsilon_0} \cos\left(\frac{\phi_0}{f}\right) \, .
\end{eqnarray}

We are interested in the $n$-point correlation functions of the variable
\begin{equation}
 \zeta(t,{\bf x}) = -\frac{H}{\dot{\phi}} \, \delta\varphi(t, {\bf x}) \, .
\end{equation}
These may be computed using the in-in formula
\begin{eqnarray}
   && \langle \zeta_{\bf k_1} \zeta_{\bf k_2} \cdots \zeta_{\bf k_n}(\tau) \rangle 
               = \sum_{N=0}^\infty (-i)^N \int_0^t dt_1 \int_0^{t_1}dt_2 \cdots \int_0^{t_{N-1}} dt_N \nonumber \\
  && \times \langle \left[ \left[ \left[ \zeta_{\bf k_1} \zeta_{\bf k_2} \cdots \zeta_{\bf k_n}(\tau) , H_I(t_1)  \right] , H_I(t_2) \right] \cdots , H_I(t_N) \right]  \rangle \, , \label{in-in}
\end{eqnarray}
where $H_I(t) = -\int d^3 x \mathcal{L}_I$ is the interaction Hamiltonian and, with some abuse of notation, it is understood that the fluctuation modes on the right hand side correspond to the free theory solutions.

We choose the free theory to correspond to the quadratic action for $\zeta$, neglecting resonance effects (setting $b=0$).  To leading order in slow roll parameters, the free theory Lagrangian is given by
\begin{equation}
\label{Lfree}
 \mathcal{L}_0 = M_p^2 a^3 \epsilon_0 \left[ \dot{\zeta}^2 - \frac{1}{a^2} (\vec{\nabla}\zeta)^2  \right] \, .
\end{equation}
The solutions of (\ref{Lfree}) are well-known.  The q-number Fourier mode is written as
\begin{equation}
 \zeta_{\bf k}(\tau) = b({\bf k}) u_k(\tau) + b^\dagger(-{\bf k}) u_k^\star(\tau) \, ,
\end{equation}
where the annihilation/creation operators are
\begin{equation}
 \left[b({\bf k}), b^\dagger({\bf k'})\right] = \delta^{(3)}\left({\bf k}-{\bf k'}\right) \, ,
\end{equation}
and the c-number modes are given by
\begin{equation}
\label{ufree}
 u_k(\tau) \approx \frac{H^2}{\dot{\phi}_0} \frac{e^{-ik\tau}}{\sqrt{2k^3}} (1 + ik\tau) \, ,
\end{equation}
up to an irrelevant overall phase.

Now we turn our attention to the leading interaction terms.  These were derived in Appendix B.  The result is
\begin{equation}
 H_I(t) \cong -\int d^3 x \left[ 
\frac{\xi}{4}\, \zeta \,\eta^{\mu\nu\alpha\beta} F_{\mu\nu} F_{\alpha\beta} + 
M_p^2 a^3 \epsilon_1 \left( \dot{\zeta}^2 - \frac{1}{a^2}(\vec{\nabla}\zeta)^2  \right) 
                + \frac{a^3}{6} (2\epsilon_0)^{3/2} \Lambda^4 \frac{M_p^3}{f^3}\sin\left(\frac{\phi_0(t)}{f}\right) \,  \zeta^3  \right] \, . \label{HItext}
\end{equation}
This result is derived neglecting inhomogeneities of the metric, which has been show to be a good approximation both in the case where inverse decay effects dominate \cite{ai2}, and also when resonance effects dominate \cite{leblond}.  The first term in (\ref{HItext}) leads to inverse decay effects.  The contribution to the 2-point and 3-point correlation functions from this term is studied in Appendix C.  There we show explicitly that the in-in formula (\ref{in-in}) gives the same result as the method employed in \cite{ai,ai2}.  The last two terms in (\ref{HItext}), on the other hand, lead to resonance effects .  The relevant contributions to the 2-point and 3-point functions are computed in Appendix D.  Here we notice these two distinct physical effects ``add up'' in a very simple way, at the level of the $n$-point correlation functions.

\subsection{The Spectrum and Bispectrum}
\label{subsec:n-pt}

The 2-point correlation function defines the power spectrum
\begin{equation}
  \langle \zeta_{\bf k} \zeta_{\bf k'}(\tau) \rangle = \frac{2\pi^2}{k^3} \, P_\zeta(k) \, \delta^{(3)}\left({\bf k}+{\bf k'}\right) \, .
\end{equation}
The $N=0$ term in (\ref{in-in}) gives the usual (nearly) scale-invariant result.  There are two important corrections to this.  At $N=1$ there is a contribution due to the quadratic interaction terms in (\ref{HItext}); these gives an oscillatory modulation of the power spectrum that was accounted for in \cite{monodromy2,pajer} using a different approach.   See Appendix D for a derivation.  At $N=2$ there is an important contribution from the pseudo-scalar coupling which influences the normalization of the spectrum; this was accounted for in \cite{ai,ai2} and is re-derived in Appendix C.  Accounting for all important contributions, we can write the power spectrum in the form
\begin{equation}
\label{Ptot}
  P_\zeta(k) = \Delta^2 \left(\frac{k}{k_\star}\right)^{n_s-1 + \frac{\delta n_s}{\ln (k/k_\star)}\cos\left(\frac{\phi_k}{f}\right)} \, .
\end{equation}
This is the same form that was analyzed in \cite{monodromy2,pajer}, however, the parameters now have the meaning
\begin{eqnarray}
  \Delta^2 &=& \mathcal{P} \left[1 +\mathcal{P} f_2(\xi) e^{4\pi\xi} \right] \, , \label{Deltasquare} \\
  \delta n_s &=&  3 b_\star \left(\frac{2\pi f}{\sqrt{2\epsilon_\star} M_p}\right)^{1/2} \left[ 1 + \mathcal{P} f_2(\xi) e^{4\pi\xi} \right]^{-1} \, . \label{Deltans}
\end{eqnarray}
Recall that $\mathcal{P}^{1/2} \equiv H^2 / (2\pi\dot{\phi})$.  The function $f_2(\xi)$ was computed in \cite{ai2} and is given by
\begin{equation}
  f_2(\xi) = \left\{ \begin{array}{ll}
         3\cdot 10^{-5} \, \xi^{-5.4} & \mbox{if $2\lsim \xi \lsim 3$};\\
         7.5\cdot 10^{-5} \, \xi^{-6} & \mbox{if $\xi \gg 1$}.\end{array} \right.
\end{equation}
(this function is plotted in Figure 1 of \cite{ai2}).

The total bispectrum is defined by
\begin{equation}
 \langle \zeta_{\bf k_1}\zeta_{\bf k_2}\zeta_{\bf k_3}\rangle  = 
  B\left(k_i\right) \delta^{(3)}\left({\bf k_1}+{\bf k_2}+{\bf k_3}\right) \, .
\end{equation}
This quantity receives two important contributions.  At $N=1$ we have a contribution due to the cubic self-interaction term in (\ref{HItext}); this gives a resonant contribution that was studied in \cite{monodromy2,pajer,leblond}.  See Appendix D for more details.  At $N=2$ we have unimportant corrections to this result.  The leading contribution from the pseudo-scalar interaction in (\ref{HItext}) arises at $N=3$ and gives an equilateral contribution to the bispectrum that was first derived in \cite{ai}; see Appendix C.  Summing up all important contributions, leads to
\begin{eqnarray}
  B(k_i) &=& B_{\mathrm{inv.dec}}(k_i) + B_{\mathrm{res}}(k_i) \, , \nonumber \\
&=& \frac{9(2\pi)^{5/2}}{10} P_\zeta^2(k) \, f_{NL}^{\mathrm{inv.dec}}(\xi) \, T_{\mathrm{eq}}(k_i)
\nonumber \\
  && +  (2\pi)^{5/2} P_\zeta^2(k) \, \frac{f_{NL}^{\mathrm{res}}}{k_1^2k_2^2k_3^2} \, \left[ 
-  \sin\left(\frac{\phi_K}{f}\right) 
  + \frac{f}{\sqrt{2\epsilon_\star}M_p} \cos\left(\frac{\phi_K}{f}\right)  \sum_{i\not= j}\frac{k_i}{k_j}   \right] \, ,
\end{eqnarray}
where the nonlinearity parameters are
\begin{eqnarray}
  f_{NL}^{\mathrm{inv.dec}} &=&  \frac{f_3(\xi) \mathcal{P}^3 e^{6\pi\xi}}{P_\zeta^2(k)} \, , \\
   f_{NL}^{\mathrm{res}} &=& \frac{3 b_\star \sqrt{2\pi}}{8} \left(\frac{\sqrt{2\epsilon_\star}M_p}{f}\right)^{3/2} \, .
\end{eqnarray}
The equilateral shape template is given by
\begin{equation}
\label{Tequil}
  T_{\mathrm{eq}}(k_i) \equiv -\frac{1}{k_1^3k_2^3}-\frac{1}{k_1^3k_3^3}-\frac{1}{k_2^3k_3^3} - \frac{2}{k_1^2k_2^2k_3^2}  + \frac{1}{k_1k_2^2k_3^3} + (5-\mathrm{perms})
\end{equation}
and the function $f_3(\xi)$ is
\begin{equation}
 f_3(\xi) = \left\{ \begin{array}{ll}
         7.4\cdot 10^{-8} \, \xi^{-8.1} & \mbox{if $2\lsim \xi \lsim 3$};\\
         2.8\cdot 10^{-7} \, \xi^{-9} & \mbox{if $\xi \gg 1$}.\end{array} \right. \label{f3}
\end{equation}
(this function is plotted in Figure 2 of \cite{ai2}).

To be precise, the shape of the non-gaussianity from inverse decay does not coincide exactly with the equilateral template (see Figure 6 of \cite{ai2}). The cosine (defined in \cite{shape}) between the precise shape and the equilateral template  evaluates to 0.94 \cite{ai2}, so that the equilateral shape can be used as a good estimate of the actual result. A positive detection of non-gaussianity based on the equilateral template will justify a more precise analysis where the exact inverse decay shape is taken into account.

The results of this subsection illustrate two important points.  First, we have seen explicitly that the in-in method reproduces the same answer as the formalism developed in \cite{ai,ai2} for inverse decay effects.  This serves as a useful consistency check and helps to verify the equivalence of two of the most popular methods for computing cosmological correlation functions.  Second, we have shown that inverse decay effects and resonance effects ``add up'' in a simple way, both at the level of the spectrum and at the level of the bispectrum.

It is interesting to note that the two different effects which we have discussed -- resonance and inverse decay -- lead to very different patterns of higher order correlation functions \cite{sarah}.  This difference may be relevant for observables that are sensitive to the pattern of moments, such as the statistics of rare objects or Minkowski functionals.


\subsection{Observational Consequences at large scales}
\label{subsec:observables}

Models of axion inflation have a rich phenomenology. The presence of a periodic modulation in the potential (\ref{V}) gives rise to resonant effects, while the inflaton coupling to gauge fields results in inflaton perturbations from the inverse decay of the produced quanta. The resonant signatures were studied in ref. \cite{chen2,monodromy2,pajer,leblond,monodromyNG} while the inverse decay signatures were studied in \cite{ai,ai2}. For definiteness, we discuss the resonant effects only for the linear case here,  $\Vsr \cong \mu^3 \varphi$ during inflation.  (This case can be derived from string theory \cite{monodromy,monodromy2} and will be discussed in more detail in sections \ref{s:review} and \ref{s:modin}.) The parameter $\mu$ is fixed by matching the observed power spectrum normalization.  There remain 3 key parameters which determine the cosmological observables: the axion scale $f$, the amplitude $\Lambda$ of the periodic term in the inflaton potential (equivalently, we can use (\ref{b}) to replace $\Lambda$ with $b$), and the coupling of the inflaton to gauge fields $\alpha$.  At a given choice of $f$, the magnitude of resonance effects are controlled by $b$ and the magnitude of inverse decay effects are controlled by $\alpha$.

\begin{figure}[h]
\begin{center}\includegraphics[width=0.5\textwidth]{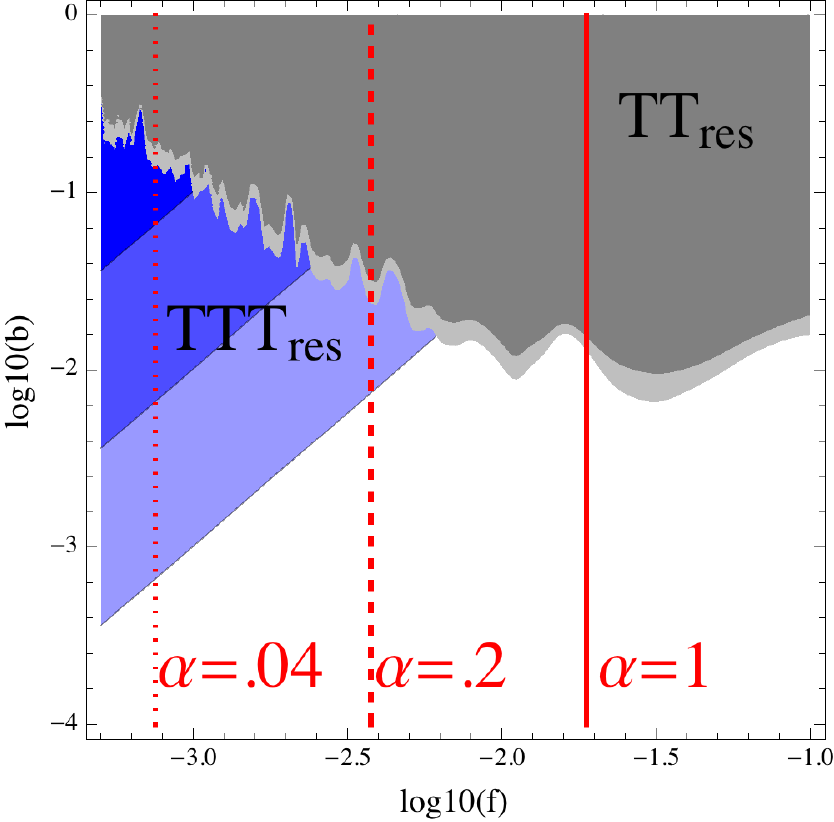}\end{center}
\caption{Parameter space for axion monodromy inflation with $\Vsr \cong \mu^3 \varphi$.  The ``TT(res)'' regions give the (one- and two-sigma) likelihood contours for the temperature 2-point function.  The ``TTT(res)'' regions give $f_{\mathrm{res}}=200,20,2$.  The region to the left of each horizontal line is ruled out by observational constraints on $f_{NL}^{\mathrm{equil}}$, for values of the coupling $\alpha=0.04,0.2,1$ (see eq. (\ref{Lint})).  The Figure has been adapted from  \cite{monodromy2} by adding the exclusion regions  from  $f_{NL}^{\mathrm{equil}}$.
}
\label{Fig:param}
\end{figure}

The phenomenological constrains on these parameters  are summarized in Figure \ref{Fig:param}.  The ``TT(res)'' regions show one- and two-sigma likelihood contours for the power spectrum, with the darker regions at large $b$ being disfavoured by the too large oscillations  in the power spectrum induce by resonant effect. The ``TTT(res)'' regions give the value of resonant nongaussianity with (from top left to bottom right) $f_{\mathrm{res}} > 200, 20, 2$, respectively. The exact bounds on the size of resonant non-Gaussianity from current and future experiments are not yet known, but research in this direction is in progress. The horizontal lines indicate the value of equilateral nongaussianity due to inverse decay.  For a given choice of $\alpha$, the region to the left of the horizontal red line is ruled out by the 95 $\%$ C.L. WMAP \cite{wmap7} bound  $f_{NL}^{\mathrm{equil}} < 266$. Recall that the generic expectation from effective field theory is $\alpha=\mathcal{O}(1)$.  We see that, in this case, observational constraints on equilateral non-Gaussianity can rule out much of the parameter space where resonant non-Gaussianity is large, but still allow for detectable oscillations in the power spectrum.  To have a reasonably large resonant bispectrum without a large inverse decay signal requires $\alpha$ much smaller than $\mathcal{O}(1)$.  For such values of the pseudo-scalar coupling, we may have a richness of possible signature in the CMB with oscillations in the spectrum and bispectrum, along with equilateral non-Gaussianity.  

\begin{figure}[h]
\includegraphics[width=0.47\textwidth]{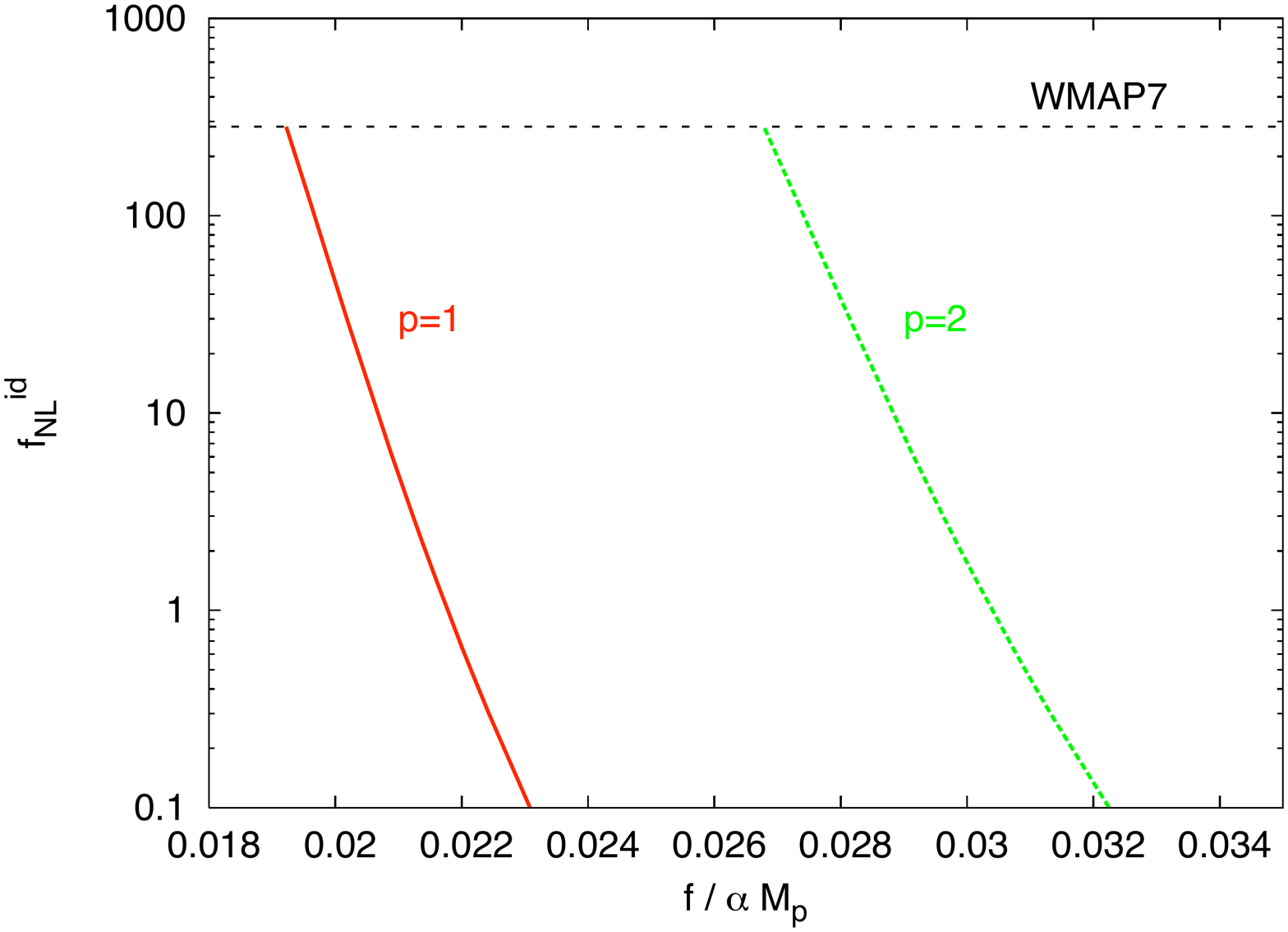}
\includegraphics[width=0.47\textwidth]{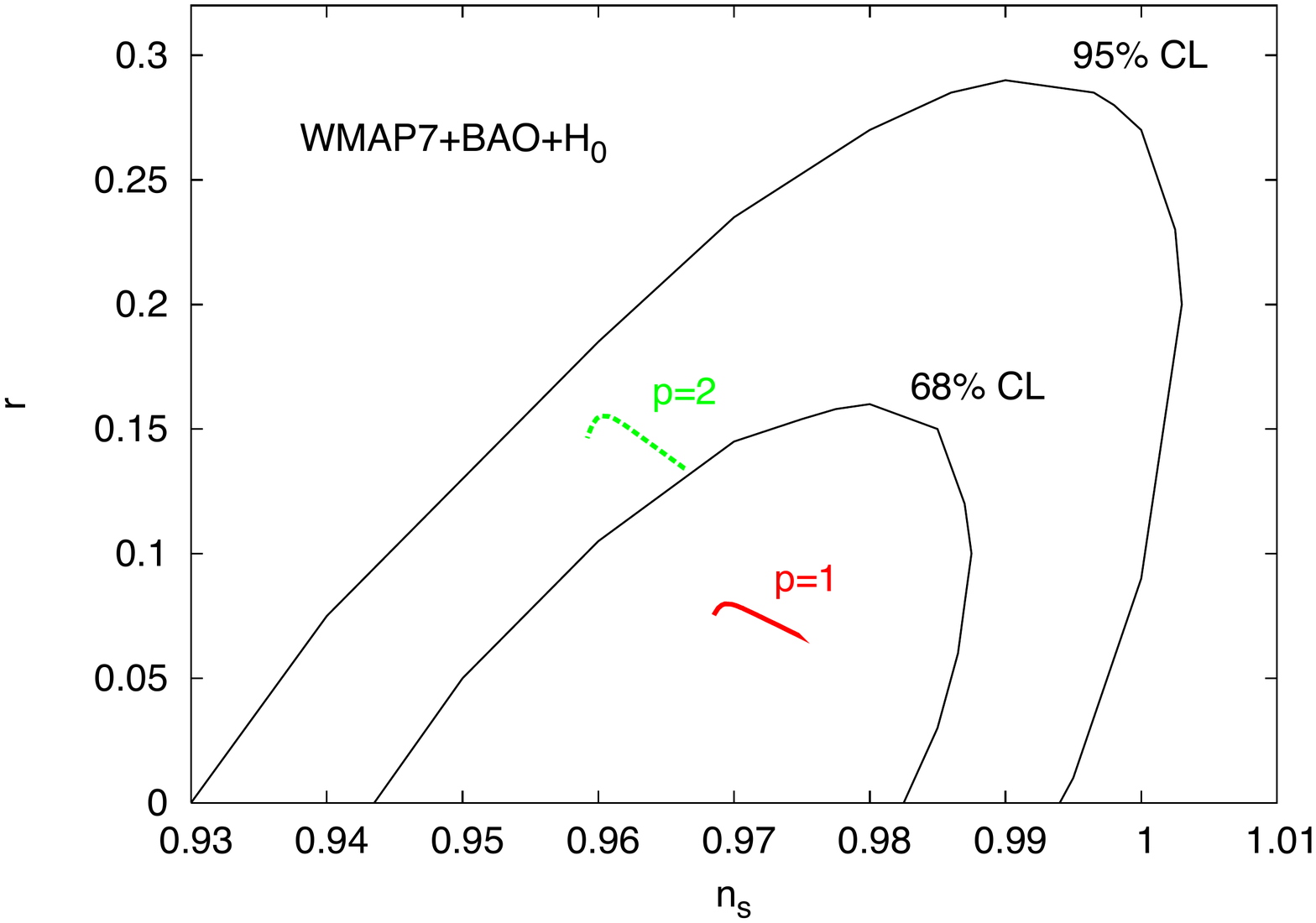}
\caption{The left panel shows the inverse decay non-gaussianity as a function of the inverse of the inflaton-gauge field coupling (see eq. (\ref{Lint})). The value shown corresponds to modes that left the horizon $60$ e-foldings before the end of inflation. The WMAP limit is obtained from the  95 $\%$ C.L. WMAP \cite{wmap7} bound  on equilateral nongaussianity, and from the fact that the shape of inverse decay non-gaussianity has a cosine of $0.94$ with the equilateral template \cite{ai2}. The right panel shows values of $n_s$ and $r$ for the same range of inflaton-gauge field couplings. These values are compared with the  68 $\%$ C.L. and  95 $\%$ C.L. limits given in \cite{wmap7}. Both panels show results for linear ($p=1$) and quadratic ($p=2$) slow-roll inflaton potentials.\label{Fig:ns-r}}
\end{figure}

In Figures \ref{Fig:ns-r} we present instead results both for linear and quadratic $\Vsr$ during inflation. In the left panel we show the value of inverse decay non gaussianity as a function of $f/\alpha$. We recall that the inflaton-gauge field coupling is inversely proportional to $f/\alpha$, see eq. (\ref{Lint}), which explains why the resonant decay non-gaussianity is a decreasing function of this quantity. The rapid decrease observed in the Figure is due to the fact that the effect is exponentially sensitive to the coupling \cite{ai,ai2}. In the right panel we show instead the phenomenological predictions for the spectral tilt $n_s$ and the tensor to scalar ratio $r$ as the inflaton-axion coupling ranges in the values shown in Figure  \ref{Fig:ns-r}. The predictions are compared with the  68 $\%$ C.L. and  95 $\%$ C.L. limits given in \cite{wmap7}.

The results shown in Figure \ref{Fig:ns-r} improve over the existing literature \cite{ai2} by taking into account the backreaction effects of the produced quanta on the background evolution.
As we have discussed in Subsection \ref{subsec:classicalevolution},  for inflaton-gauge field coupling  leading to interesting non-gaussianity, backreaction is negligible when the scales relevant for CMB (and LSS) observations left the horizon. However, it becomes relevant at the last stages of inflation. The main backreaction effect is the slow down of the inflaton due to its loss of kinetic energy into gauge quanta. This in turns increases the duration of inflation with respect to the case of no production. As a consequence, the number of e-foldings $N_{CMB}$  inside inflation at which  any CMB scale left the horizon is a function of both the initial value of the inflaton $\phi_{CMB}$ and of the coupling of the inflaton to gauge fields $\alpha / f$. If we fix the value of $N_{CMB}$ associated to a given scale (this request can be done once we know the reheating history after inflation), the increased of the duration of inflation  needs to be compensated by a value of $\phi_{CMB}$ which would lead to a smaller value of $N_{CMB}$ in the absence of the coupling. For the monomial slow roll potentials considered here, this means a smaller value of $\phi_{CMB}$. Therefore, Figure \ref{Fig:ns-r} has been obtained for a progressively smaller value of $\phi_{CMB}$ as the coupling $\alpha / f$ increases.

The bottom-right end of  each of the two lines shown in the right panel of Figure \ref{Fig:ns-r} corresponds to the standard chaotic inflation results obtained in absence of coupling. Such values are obtained as long as roughly  $\xi_{CMB} \lsim 1$ (corresponding to $f/\alpha  \gsim 0.046 M_p$ in the linear case, and  $f/\alpha  \gsim 0.065  M_p$ in the quadratic case). The upper-left ends of the two curves correspond instead to the maximal couplings compatible with the CMB non-gaussianity limit (namely, we exclude $\xi_{CMB} \gsim 2.66$,  corresponding to $f/\alpha  \lsim 0.019 M_p$ in the linear case, and  $f/\alpha  \lsim 0.027  M_p$ in the quadratic case). As the coupling increases towards the maximal allowed value, the corresponding decrease in $\phi_{CMB}$ implies a less flat potential when the CMB modes leave the horizon. This in turns results in increased values for $1-n_s$ and $r$ (this is the same reason behind the lines in the $n_s - r$ plot as a function of the number of e-foldings in the standard case, cf. Figure 20 of \cite{wmap7}). In the present case, there is an additional effect on $r$ at large couplings. Namely, the gauge quanta give a greater contribution to the scalar power spectrum than to the tensor power spectrum \cite{ai,ai2}.
Specifically, the standard  relation $r \simeq 16 \, \epsilon$ is modified into $r \simeq 7 \epsilon^2$ in the limit in which the inverse decay perturbations dominate over the vacuum one \cite{ai2}. As shown in \cite{ai,ai2}, the inverse decay  non-gaussianity limit forces one to be in the regime in which the power is dominated by the vacuum mode. For the maximal possible couplings, the inflaton perturbations from gauge modes contribute to about $10 \%$ of the scalar power spectrum, while the corresponding contribution to the tensor power is about $3 \cdot 10^{-4}$ for linear inflaton potential, and about
 $6 \cdot 10^{-4}$ for quadratic inflaton potential. This is the cause of  the decrease of $r$ at the left end of the curves shown in the right panel of Figure \ref{Fig:ns-r}.


\section{Gravitational Waves at Interferometers}
\label{sec:LIGO}

In Section \ref{sec:CMB} we discussed the observable cosmological fluctuations on CMB/LSS scales.  Such scales left the horizon roughly $55$ to $60$ e-foldings before the end of inflation, during the phase where backreaction effects are negligible.  In this section, we instead study scalar and tensor fluctuations on much smaller scales.  These modes left the horizon closer to the end of inflation, when backreaction effects start to play an important role in determining the evolution of the homogeneous background, $\phi(t)$ and $H(t)$. Our main results are summarized in figure \ref{GW2}, where we show that   Advanced LIGO/VIRGO could detect a stochastic background of gravitational waves from inflation for $\xi_{CMB}$ as small as $2.33$  (equivalent to $f/(\Mpl \alpha)\leq 0.021$) in the case of a linear inflaton potential, and as small as $2.23$  (equivalent to $f/(\Mpl \alpha)\leq 0.031$) in the case of a quadratic potential.

\begin{figure}
\includegraphics[width=.5\textwidth]{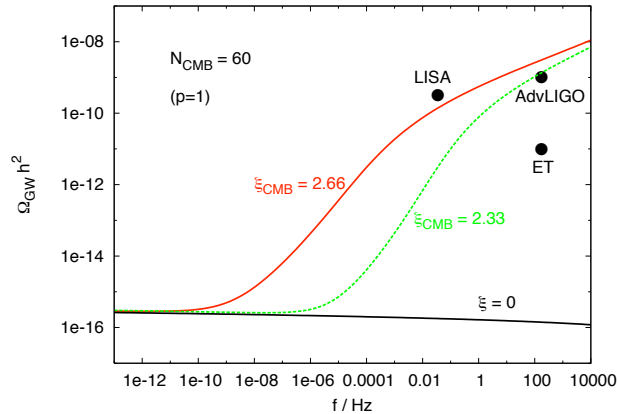}
\caption{$\Omega_{GW} \, h^{2}$ as function of the frequency $f$, for $N=60$ e-foldings of observable inflation,  a linear slow roll inflaton potential,  and $\xi_{CMB}=0,\,2.33,\,2.66$ (the value of $\xi$ when the large scale CMB modes left the horizon). For reference we also show the expected sensitivity of LISA, Advanced LIGO/VIRGO and Einstein Telescope (at their most sensitive frequency). \label{GW}}
\end{figure}

\begin{figure}
\includegraphics[width=.45\textwidth]{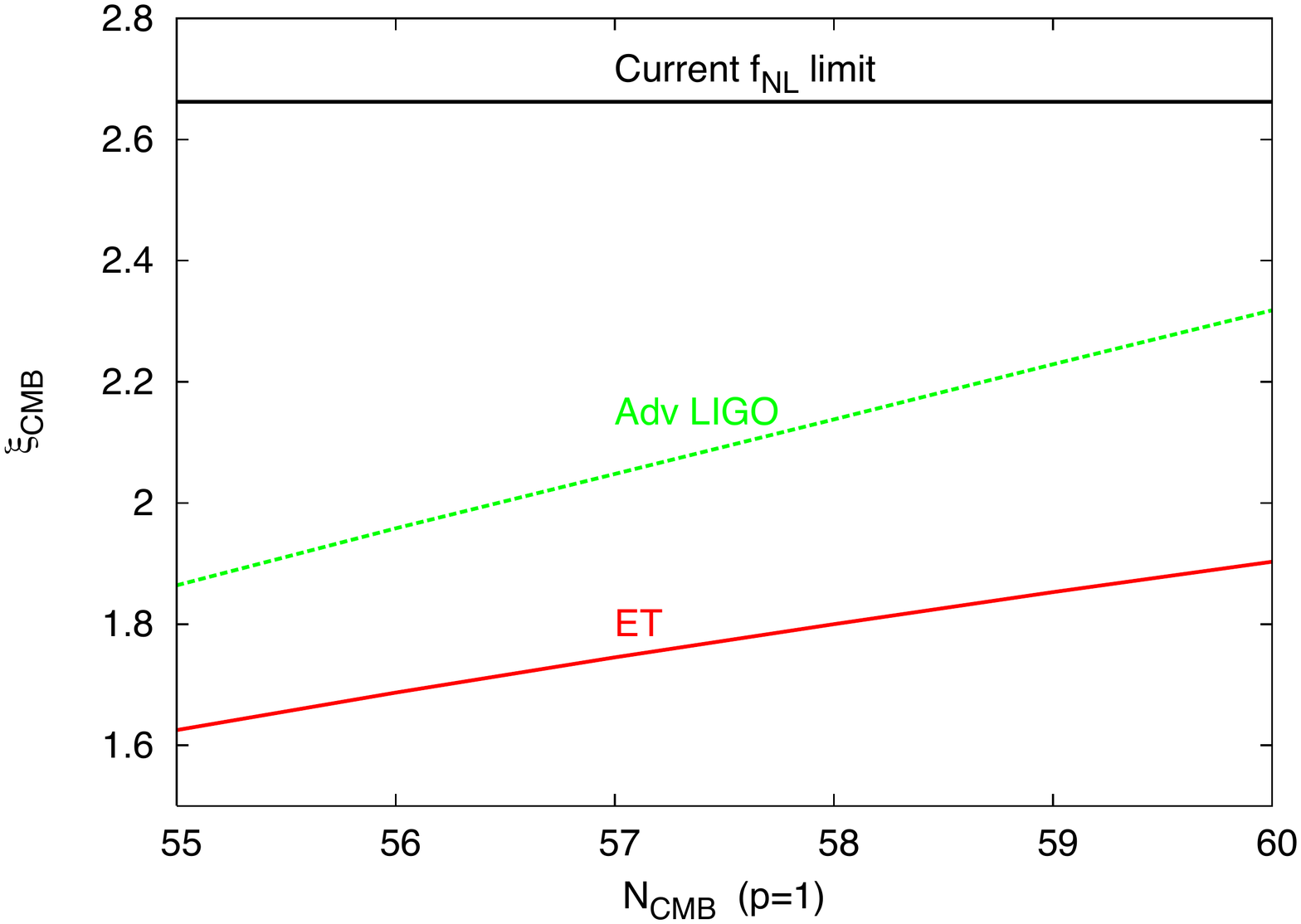}
\includegraphics[width=.45\textwidth]{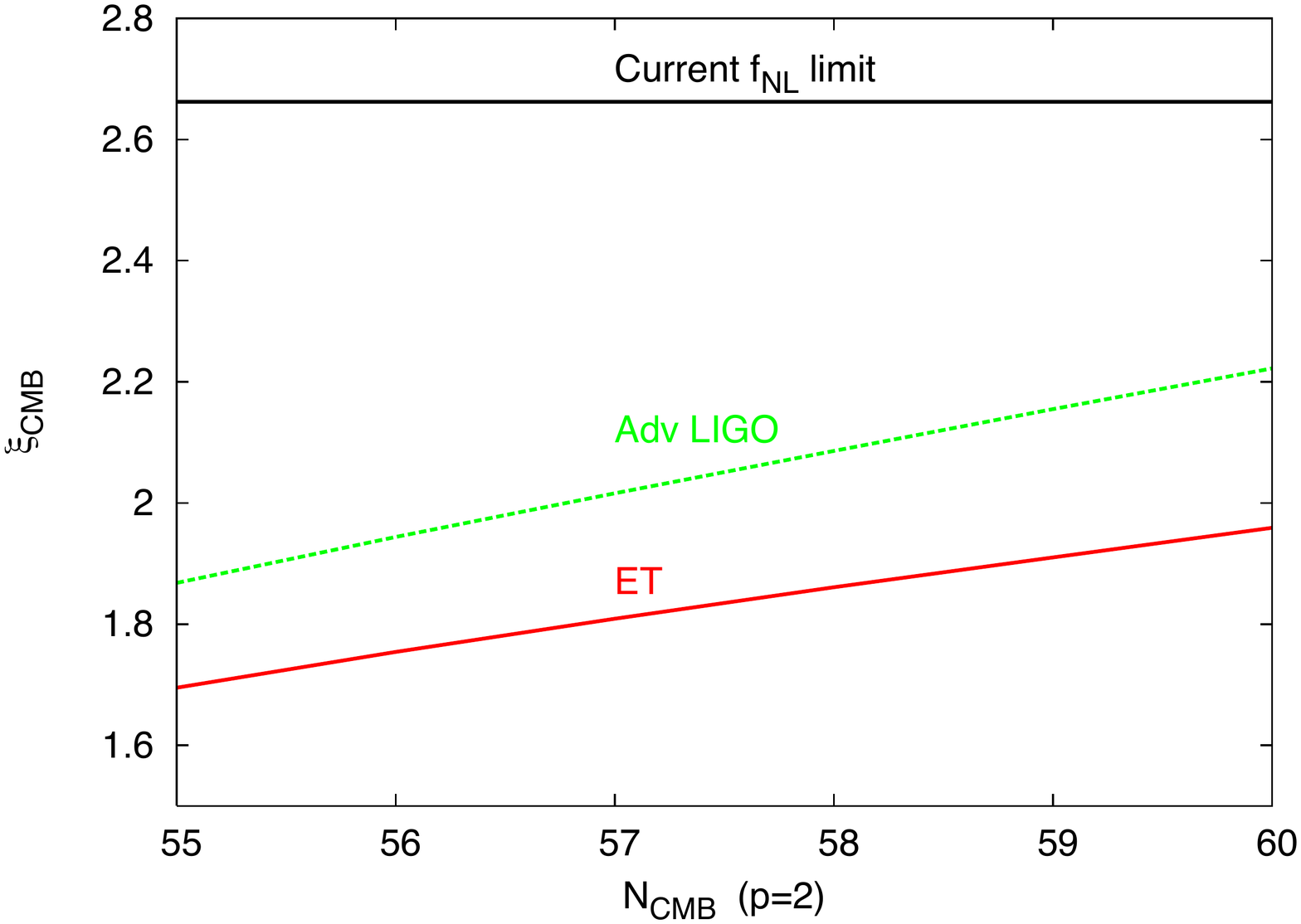}
\caption{Region in the $\{N_{CMB},\xi_{CMB}\}$ plane (values assumed by these quantities when the large scale CMB modes left the horizon) for which  the gravity wave  signal is detectable at Advanced LIGO/VIRGO and Einstein Telescope. The left and right panel refer to a linear and quadratic inflaton potential, respectively.\label{GW2}}
\end{figure}


\subsection{The Gravitational Wave Signal}
\label{subsec:GW}

The inverse decay of tachyonic gauge fields does not only affect scalar but also tensor perturbations. The power spectrum has been computed in \cite{ai}, and further studied in \cite{parity,ai2}.
Defining $g_{ij}=a^{2}(\delta_{ij}+h_{ij})$ with $h_{ij}$ transverse and traceless, $h_{ii}=\partial_{i}h_{ij}=0$, and its Fourier transform as
\be
h_{ij}(x,t)=\int \frac{d^{3}k}{(2\pi)^{3/2}}\sum_{r=L/R}\Pi_{ij,r}(\vec k) \left[h_{r}(k,t)a_{\vec k}+h.c.\right]
\ee
one finds the tensor power spectra
\be\label{Ph}
P_{L/R}&\equiv& \frac{k^{3}}{2\pi^{2}}|h_{L/R}|^{2}\nonumber \\
&=&\frac{H^{2}}{\pi^{2}\Mpl^{2}}\left(\frac{k}{k_{p}}\right)^{n_{T}}\left[1+\frac{2H^{2}}{\Mpl^{2}}f_{h,L/R}(\xi)e^{4\pi \xi}\right]\,,
\ee
for the left and right polarization respectively. Here the functions $f_{h,L/R}$ are defined by
\begin{eqnarray}
f_{h,L/R} & = & \frac{1}{\xi}   \, \int \frac{d^3 q_*}{\left( 2 \pi \right)^3} \,  \sqrt{  q_*  \, \vert {\hat k} - {\bf q_*} \vert } \;  \frac{ \left( 1 \pm \cos \theta \right)^2 \left( 1 - q_* \cos \theta \pm \sqrt{1 - 2 q_* \cos \theta + q_*^2} \right)^2 }{ 16 \left( 1 - 2 q_* \cos \theta + q_*^2 \right)} \times  \nonumber\\
  &&  \!\!\!\!\!\!  \left\{ \int_0^\infty d x \sqrt{  x }  \left[  \sin x - x \, \cos x  \right] \,
\left[ \frac{2 \xi}{x} +  \sqrt{ q_*  \, \vert {\hat k} - {\bf q_*}  \vert } \right]
 {\rm e}^{-2 \sqrt{2 \xi x} \left[ \sqrt{ \vec{q}_* } + \sqrt{\vert {\hat k} - \vec{q}_* \vert} \right]} \right\}^2
\end{eqnarray}
where $q_*$ and $\theta$ are, respectively, the magnitude of the  (dimensionless) integration momentum ${\bf q}_*$, and the angle between this vector and the momentum ${\bf k}$ of the mode.
For details, and for the numerical evaluation of these functions, see \cite{ai2}.

In order to compare our results with the sensitivity of interferometers, it is convenient to introduce the fractional energy density per logarithmic wavenumber interval
\begin{eqnarray}
\Omega_{GW}\equiv\frac{1}{3H_{0}^{2}\Mpl^{2}}\frac{\partial \rho_{GW,0}}{\partial \log k}\,,
\end{eqnarray}
where the label $0$ refers to quantities evaluated nowadays and
\begin{eqnarray}\label{rho}
\rho_{GW,0}\equiv\frac{\Mpl^{2}}{8a^{2}}\langle h_{ij}'h_{ij}'+\partial_{k}h_{ij}\partial_{k}h_{ij} \rangle\,.
\end{eqnarray}
To evaluate $\rho_{GW,0}$ we need to evolve the primordial signal \eqref{Ph} after horizon re-entering until today. Using the tensor transfer function (see e.g.~\cite{Weinberg,Boyle:2005se}) for modes that re-entered the horizon during radiation domination, as it is the case for the modes relevant for interferometers, one finds
\begin{eqnarray}
\Omega_{GW}=\frac{\Omega_{R,0}}{24}\left(P_{L}+P_{R}\right)\,,
\end{eqnarray}
where $\Omega_{R,0}\equiv \rho_{R,0}/3H_{0}\Mpl^{2}\simeq 8.6\cdot 10^{-5}$ refers to radiation nowadays (one needs to include also the neutrinos as if they were still relativistic today). It is useful to plot $\Omega_{GW,0}$ as a function of frequency $f=k/2\pi$, which is related to the number of efoldings by
\be
N-N_{CMB}&=&\ln \frac{k_{CMB}}{0.002 {\rm Mpc}^{-1}}-\ln \frac{k}{0.002 {\rm Mpc}^{-1}}\nonumber\\
&=&\ln \frac{k_{CMB}}{0.002 {\rm Mpc}^{-1}}-44.9-\ln \frac{f}{10^{2}{\rm Hz}}\,.
\ee

The resulting signal as a function of frequency is presented in  Figure \ref{GW} for the case of a linear inflaton potential - which is an excellent approximation to the potential for axion monodromy \cite{monodromy} -  and for three reference values of the inflaton-gauge field coupling. Specifically we choose $\xi_{CMB} = 2.66$, corresponding to  roughly  the highest value allowed by the CMB limits on equilateral non-gaussianity,  $\xi_{CMB} = 2.33$, corresponding to roughly  the lowest possible value that can be detected at Advanced LIGO/VIRGO, and $\xi_{CMB} = 0$, corresponding to standard vacuum fluctuations.

The results presented in the left panel of Figure \ref{GW} have been obtained assuming  $N_{CMB}=60$. A lower duration of inflation leads to a greater signal, as we show in the right panel\footnote{As we discussed in subsection \ref{subsec:reheating}, for $\xi = {\rm O } \left( 1 \right)$, having $N_{CMB} < 60$ requires however a non-minimal reheating history.}. The fact that the signal grows with $\xi_{CMB}$ is due to the exponential term in \eqref{Ph}, while its growth with decreasing $N_{CMB}$ is slightly more subtle. For fixed $\xi_{CMB}$, decreasing $N_{CMB}$ affects the scales relevant for interferometers ($N=N_{CMB}-44.9$) in two ways: it decreases $H$, since these scales are now closer to the end of inflation and it increases $\xi$, since the $\dot \phi/H$ is larger. The latter effect ends up increasing the total gravitational signal due to the exponential enhancement. 

Finally, let us comment on how the consistent account of the strong backreaction phase affects the results of \cite{GWlorenzo}, where this phase was neglected. One effect can be seen in Figure \ref{GW} for a fixed $\xi_{CMB}\neq0$, where one can distinguish three phases. From left to right one sees the vacuum contribution, then the fast growth of the inverse decay contribution and finally the slow down of the growth when the backreaction becomes sizable. A second effect, which is clearly visible by comparing the two curves for different $\xi_{CMB}$ is that for larger $\xi_{CMB}$ the whole signal is effectively shifted toward lower frequencies (to the left) due to the larger number of additional efoldings of strong backreaction at the end of inflation. The two effects act in opposite direction, but since the latter is stronger than the former, the net result of the strong backreaction phase is an \textit{increase} of the signal. This increase is crucial in order to make the natural window $N_{CMB}\sim 59-62$ potentially visible at Advanced LIGO/VIRGO.


\subsection{Perturbations in the Regime of Strong Backreaction}

So far, we have discussed the tensor perturbations on small scales, relevant for interferometers.  One should also study  scalar cosmological fluctuations at such scales.  Although these are not directly observable (besides the possible generation of primordial black holes), it is nevertheless important to verify that scalar fluctuations are perturbatively small, as a consistency check.  For scales which left the horizon $\sim 10-20$ e-foldings before the end of inflation, the calculation presented in Section \ref{sec:CMB} is not valid, since backreaction effects no longer have a negligible impact on the homogeneous dynamics.  Let us provide a brief, heuristic analysis of the scalar fluctuations in the regime of strong backreaction, in order to justify the validity of perturbation theory (which was implicit in our previous analysis of gravitational waves).  According to Ref.~\cite{lorenzo}, the inflaton perturbations obey an equation of the form

\begin{equation}
\label{dissipative_eqn}
 \delta\ddot{\varphi} + 3 \beta H \delta\dot{\varphi} - \frac{\vec{\nabla}^2}{a^2} \delta\varphi + m^2 \delta \varphi = J  \, ,
\end{equation}
where the effective coefficient of friction, $\beta$, and the source term, $J$, are given by

\begin{equation}
 \beta \equiv 1-2\pi\xi\frac{\alpha}{f} \frac{\langle \vec{E}\cdot\vec{B}\rangle}{3 H \dot{\phi}} \, , \hspace{5mm}
 J = \frac{\alpha}{f}\left[ \vec{E}\cdot\vec{B} - \langle \vec{E}\cdot\vec{B}\rangle \right]  \, .
\end{equation}
We have assumed that $\xi,\dot{\phi} > 0$ and (\ref{backeqns}) shows that $\langle \vec{E}\cdot\vec{B} \rangle < 0$, so the effect of particle production is always to \emph{increase} the effective friction in (\ref{dissipative_eqn}), as would be expected on physical grounds.  Equations similar to (\ref{dissipative_eqn}) have also been obtained in other scenarios where particle production effects contribute an important source of friction; see \cite{trapped,senatore_new} for example.

When backreaction effects on the homogeneous dynamics are negligible we have $\frac{\alpha}{f} |\langle\vec{E}\cdot\vec{B}\rangle| \ll 3 H |\dot{\phi}|$; see Section \ref{sec:bkg}.  In this case $\beta \cong 1$ and (\ref{dissipative_eqn}) coincides with the equation studied in \cite{ai,ai2}.  Instead, when backreaction effects are important $\beta \gg 1$ is possible.  We can give a heuristic estimate of the size of scalar fluctuations in the regime of strong backreaction as follows.  Since most of the interesting production occurs near horizon crossing, we estimate the derivatives as $\partial \sim H$.  In this case, and for $\beta\gg 1$, \eqref{dissipative_eqn} gives: $3 \beta H^2 \langle \delta\varphi^2 \rangle^{1/2} \sim \langle J^2 \rangle^{1/2}$.  We further estimate $\langle J^2 \rangle^{1/2} \sim \frac{\alpha}{f}|\langle\vec{E}\cdot\vec{B}\rangle|$ and $\zeta \sim - \frac{H}{\dot{\phi}} \delta\varphi$.  The variance of fluctuations is roughly
\begin{equation}
\label{Pdissipation}
 \langle \zeta^2 \rangle \sim \mathcal{O}(10^{-2}) \, \frac{1}{\xi^2} \, ,
\end{equation}
which is consistent with the analysis of \cite{lorenzo}.  Equation (\ref{Pdissipation}) suggests that curvature fluctuations on small scales are much larger than on CMB scales, although a perturbative analysis is still justified.  (A useful benchmark value to keep in mind is $\xi \sim 5-6$ near the end of inflation for the case where non-Gaussianity is observable, but not yet ruled out.)


\section{Axion Monodromy Inflation}\label{s:review}

We start this section by reviewing a construction of axion monodromy inflation in string theory following \cite{monodromy,monodromy2} and a generalization proposed in \cite{dante}. We then conclude by summarizing and commenting on the main open issues in the current implementation of this class of models.


\subsection{Review of the Model}

Consider an O3/O7 orientifold Calabi-Yau compactification of Type IIB string theory. The relevant four-dimensional axion\footnote{$B$-type axions, coming from the NSNS two-form, receive a large mass from the non-perturbative corrections used in the moduli stabilization \'a la KKLT. Although one might very well conceive a scenario in which moduli are stabilized by perturbative corrections, this is harder to construct in detail.} comes from the RR two-form $C_{2}$ upon dimensional reduction. Following \cite{monodromy2}, we choose the normalization
\be
\int_{\Sigma_I} \omega_J=(2 \pi)^2 \al \delta^{~I}_J
\ee
for a basis $\Sigma_I$ of the homology $H_2(Y,\mathbb{Z})$ and $\omega_{J}$ of the cohomology $H^{2}(Y,\mathbb{Z})$, where $Y$ is the Calabi-Yau space over which we compactify. With an ansatz for the ten-dimensional RR two-form
\be
C_2=\frac{1}{2\pi}c(x) \omega\,,
\ee
for some base two-cycle $\omega$, we get a four-dimensional axion $c(x)$ with periodicity\footnote{This can be seen {\it e.g.} via S-duality starting from the world-sheet coupling $\int B /(2\pi \al)$. Note that this choice differs from that in \cite{monodromy}, where the axion periodicity was $(2\pi)^2$.} $2\pi$. The axion decay constant for $c$ is\footnote{Let us check that this is compatible with e.g. (3.5) of \cite{trivedi} and (2.15) of \cite{monodromy}. The volume scaling $L^{-4}$ agrees since $v^{\alpha}\sim L_{E}^{2}$. For the scaling with $g_{s}$ one has to transform from the string-frame volume of \cite{Panda:2010uq,monodromy} to the Einstein-frame volume of \ref{f}, i.e. $L^{4}_{s}=g_{s}L^{4}_{E}$, so that $(f/\Mpl)^{2}\sim g_{s}^{2}/L^{4}_{s}=g_{s}/L_{E}^{4}$. The fact that \cite{Panda:2010uq,monodromy} use a different 4D Einstein frame from the one we use here doesn't matter since $f$ and $\Mpl$ scale the same under a constant red-shift of the metric so that the ratio is invariant.}
\be\label{f}
\frac{f^2}{\Mpl^2}=\frac{g_s}{12 \V_E (2\pi)^2} \left[\frac{\int \omega\wedge \ast \omega}{(2\pi)^6 \al^3}\right]=\frac{g_{s}}{8\pi^{2}}\frac{c_{\alpha -- }v^{\alpha}}{\V_{E}}\,,
\ee
where $\V_{E}$ is the six-volume of $Y$ measured with the Einstein frame metric. In order to break the axion shift-symmetry (the details of the resulting potential will be reviewed in section \ref{s:modin}), we consider a spacetime filling NS5-brane wrapping the cycle $\Sigma$ that defines the axion $c$. If this cycle is far away from any O-plane, the local physics is given by the oriented Type IIB string theory, since the orientifold projection relates the fields at some point $x$ to the fields at some other far away image point $x'$ and so does not lead to any local constraint. Tadpole cancellation requires an additional anti-NS5-brane wrapping a homologous two-cycle. A metastable configuration can be obtained if we assume that $\Sigma$ belongs to a one parameter family of cycles that stretches between two warped throats and brane and anti-brane are located in different throat. Since the tension of the branes is reduced by warping, there is a potential barrier for the branes to annihilate. Inflation takes place as the $C_{2}$ flux over $\Sigma$ slowly decreases to minimize the NS5-brane (an anti-brane) tension. A series of consistency checks and bounds have been discussed in \cite{monodromy,monodromy2}. 

To conclude, we want to mention a slight modification of the above construction described in \cite{dante} that alleviate some of the difficulties encounter in this model. Imagine that for some reason it is hard to compute or control the monodromy potential generated by the introduction of NS5-branes (some concrete difficulties will be discussed in the next subsection). Inflation can still take place if there is another axion (in general one expects $h_{1,1}^{-}$ axions, i.e.~as many axions as orientifold-odd two-cycles) which interacts with the one discussed above via a non-perturbative effect (which are not forbidden by any apparent symmetry). The resulting two field potential (plotted in polar coordinates) closely resemble the Inferno as described by Dante, which was taken as the name of the model. If there is a mild hierarchy between the two axion decay constants (the precise amount depends on the details of the monodromy potential) a long stage of slow-roll inflation can be obtained.


\subsection{Issues with Concrete Realizations}

Before proceeding to the study of $\phi F\tilde F$ in the string realization of axion monodromy, we would like to stress that there are still several challenges that the explicit constructions face. Although addressing these challenges is beyond the scope of the present paper, we list some of them below.

\textit{Explicit solution.} One should provide an explicit solution with the required geometry (proposals were give in \cite{monodromy}) and solve the ten-dimensional equations of motion in the presence of warping, fluxes and an NS5/anti-NS5-brane pair. Among other things, it would then be interesting to compute the exact warped axion decay constant and the masses of NS5-brane modes.

\textit{Backreaction on four-cycles.} The $C_{2}$ flux (i.e.~the inflaton) on $\Sigma$ induces and effective (anti-)D3-brane charge on the (anti-)NS5-brane, which changes the warp fact, which in turns modifies all warped volumes. Since four-cycle volumes are stabilized, this chain of interactions leads to an effective potential for the inflaton. In the single axion monodromy model this effect can be suppressed if the NS5/anti-NS5-brane pair behaves effectively as a dipole, i.e.~when the distance between them is much smaller than the distance to the closest stack of D7-branes. In Dante's Inferno this effect is smaller to begin with \cite{dante} for two reasons. First, the flux on the two-cycle wrapped by the NS5-brane is suppressed by the ratio of the two axion decay constants, which is assumed to be small, and second the inflaton direction is almost perpendicular to the direction for which the backreaction induces a potential.

\textit{Light KK-modes.} As pointed out in \cite{monodromy2}, flux along the compact direction of brane decreases the masses of the Kaluza-Klein modes. This can be understood by considering the T-dual picture where flux becomes tilting of a lower dimensional brane, which is therefore longer that in the fluxless case. In the single axion-monodromy case, these lighter masses are very close to the Hubble scale, while they can be made parametrically heavier in the Dante's Inferno scenario.

\textit{NS5/D7-brane intersection.\footnote{This issues was raised in private communications among R.~Flauger, L.~McAllister, T.~Wrase and one of the authors}} Just on topological grounds, any two-cycle should intersect its dual four-cycle at least at one point. In the Type IIB compactifications where we understand how to stabilize the moduli (based on \cite{KKLT}) some non-perturbative corrections are needed to stabilize the K\"ahler moduli (four- and two-cycles). In the single axion monodromy model \cite{monodromy} this should arise via gaugino-condensation on a stack of spacetime filling D7-branes wrapped on a four-cycle for each homology class. The effects of the resulting NS5/D7-brane intersection on both the moduli stabilization and on the inflaton potential should be studied. Perturbative K\"ahler stabilization could in principle solve this issue but it is computationally much harder to make explicit.

In the rest of the paper we will compute certain local quantities that should be rather independent on the way the above issues are addressed.


\section{Couplings from String Theory}\label{s:modin}

We devote this section to the explicit computation of the bosonic inflaton couplings induced by the presence of an NS5 and a D5-brane and their phenomenological consequences. 


\subsection{Model Independent Couplings: NS5-brane}

The action for the NS5-brane can be obtained via S-duality from that of a D5-brane as discussed in Appendix E. In the following we focus on the potential for $c$ and its couplings to the gauge field living on the world-volume of the NS5-brane. Performing an S-duality transformation on the action for D5-branes one finds
\be 
S_{NS5}^{DBI}&=&-\al \mu_{5} \int dx^{4} \sqrt{-g_{E,4}}\sqrt{e^{\phi}|\tau|^{2}g_{E,2}+(2\pi c)^{2}}\left(1+\frac{(2\pi \al)^{2}}{4 }\frac{e^{-\phi}}{|\tau|^{2}}F_{E}^{2}\right)\,,\\
S_{NS5}^{CS}&=&\frac{3}{3!}\frac{\mu_{5}(2\pi \al)^{3}}{|\tau|^{2} }\int F\wedge F\,c \,C_{0}=\frac{1}{2 (2\pi)^{2}|\tau|^{2}}\int F\wedge F\,c \,C_{0}\,,
\ee
where a subscript $E$ indicates Einstein frame and $\tau\equiv C_{0}+i e^{-\phi}$.

Defining $\tilde F^{\rho\sigma}\equiv F_{\mu\nu} \epsilon^{\mu\nu\rho\sigma}/ (2\sqrt{-g_{E}})$, one can rewrite this action using the notation of 
section \ref{s:EFT}.  Including also the axion kinetic term, we have
\be 
S=\int d^{4}x \sqrt{-g_{E}} \left[-\frac{B(\varphi)}{4}F^{\mu\nu}F_{\mu\nu}-\frac{C(\varphi)}{4} F^{\mu\nu} \tilde F_{\mu\nu} 
- \frac{1}{2}(\partial\varphi)^2-V(\varphi) \right]\,,
\ee
where the coupling functions are
\begin{eqnarray}
 \Cc(\varphi)&\equiv& \frac{C_{0}g_{s}^{2} \varphi}{(2 \pi)^{2} f} \\
  \Bb(\varphi) &\equiv& \frac{(\alpha')^2}{2\pi |\tau|^4g_s^2} \sqrt{ g_{E,2}g_{s}|\tau|^{2}+(2\pi \varphi/f)^{2} }
\end{eqnarray}
After imposing that  the potential has a vanishingly small cosmological constant, the inflaton potential takes the form
\begin{equation}
 V(\varphi) = \mu^3 \left[ \sqrt{\varphi^2 + \varphi_c^2} -\varphi_c \right]
\end{equation}
where
\be 
\varphi_c &\equiv&\Mpl \frac{g_{s}|\tau| g_{E,2} \sqrt{c_{\alpha - -}v^{\alpha}} } {(2\pi)^{2}\V_{E}}\nonumber\\
&\simeq& \Mpl \frac{1+C_{0}g_{s}}{(2\pi)^{2}\sqrt{2}}\ll \Mpl
\ee
In the last equality we assume an isotropic compactification for a rough estimate.
During inflation $\varphi \gg \varphi_c$ and the potential can be approximated as $V(\varphi) \approx \mu^3 |\varphi|$.  (Notice that, for such large field 
values, we also have $B(\varphi) \propto V(\varphi)$.)


\subsection{NS5-brane Phenomenology}

As discussed in section \ref{subsec:gaugeprodn}, the presence of a tachyonic gauge mode depends on the parameter
\be 
\xi=\sqrt{\frac \epsilon 2} \Mpl \frac{\Cc_{,\phi}}{\Bb}\,.
\ee
To estimate this we assume that 60 e-foldings before the end of inflation $\phi \sim 10 \Mpl \gg f$, hence we can neglect the first term in the square root in the definition of $\Bb$. Our estimate is therefore
\be 
\xi\simeq \frac{\Mpl^{2}}{2\phi^{2}}C_{0}g_{s}^{3}|\tau|^{2}\,.
\ee
Notice that all powers of $f$ cancel. Now the two possible regimes are $C_{0}\ll1/g_{s}$ or the opposite $C_{0}\gg 1/g_{s}$. In each case one finds 
\be 
\xi &\simeq& \frac{\Mpl^{2}}{2\phi^{2}}C_{0}g_{s}\sim C_{0}g_{s} 10^{-2}\ll10^{-2}\quad \mathrm{or}\\
\xi &\simeq& \frac{\Mpl^{2}}{2\phi^{2}}\left(C_{0}g_{s}\right)^{3}\sim \left(C_{0}g_{s}\right)^{3} 10^{-2}\,,
\ee
respectively. To have nontrivial production of gauge field fluctuations requires $\xi \gsim \mathcal{O}(1)$.  Evidently, this is possible only when $C_{0} \gtrsim 7/g_{s} $, while no relevant effect is expected from the coupling to the NS5-brane degrees of freedom for $C_{0}\lesssim 7/g_{s} $. Larger values of $C_{0}$ seem less natural but not impossible. Allowing for fine tuning, $C_{0}$ can be made large, as one can see in the explicit examples of \cite{Lust:2006zg} where eq. (5.7) gives the value of $C_{0}$ at the minimum. As a figure of merit, we found that sampling random values for the fluxes considered in \cite{Lust:2006zg} about $.5\%$ of the cases had $C_{0}>70$. The phenomenology of the regime $C_{0}\gtrsim 7/g_{s}$ is completely analogous to the one we reviewed in section \ref{sec:CMB}. The inverse decay proceeds through the $CF\tilde F$ coupling, while $BF^{2}$ gives only subleading corrections. Potential observables are: non-Gaussianity in the CMB, running of the spectral index and gravitational waves at interferometers.


\subsection{Model Dependent Couplings: D5-brane}

One could assume that there is also a spacetime filling D5-brane wrapping a two-cycle\footnote{Again we assume that in case an orientifold projection has been performed, the two-cycle in question is far away from O-planes so the that the local physics is described by Type IIB string theory.} which contains the two-cycle $\Sigma$ that defines the axion $c$. This brane is not essential for the construction, but for completeness we include a study of its effects. In order for the D5-brane not to complicate the construction outlined in section \ref{s:review} and not to heavily affect the inflationary dynamics, it must be stabilized at a different location from the NS5-brane. This could arise because it wraps a distinct but overlapping cycle, or because of a particular configuration of fluxes.

In this case, a coupling of $c$ to the gauge theory living on the world-volume of the D5-brane is induced by the Chern-Simon part of the action. One finds
\be
S\supset  \mu_{5} (2\pi \al)^{3}  \frac12 \int F\wedge F \,c\,,
\ee
where of course this $F$ is different from the one living on the NS5-brane. The kinetic term for $F$ reads
\be 
-\frac{  v_{2}^{s}}{4 (2\pi)^{3}g_{s}}\int d^{4}x \sqrt{-g_{E}}  F_{E}^{2}\,,
\ee
where $v^{s}_{2}$ is the volume (measured in the string frame) in units of $\sqrt{\al}$ of the two-cycle wrapped by the D5-brane. Canonically normalizing one obtains
\be
S=\int d^{4}x\sqrt{-g_{E}} \left[-\frac14 F^{2}_{E}-\frac{\alpha \phi}{4 f}F_E\tilde{F}_E\right]\,,
\ee
with\footnote{Let us compare this with (6.7) of \cite{Panda:2010uq}. They define $L^{2}$ as the string-frame volume measured in units of $\sqrt{2\pi \al}$ as can be seen from their (6.5)-(6.6). This accounts for the $2\pi$ difference. The factor of $g_{s}$ on the other hand is mysterious for me. They should have a $g_{s}^{-1}$ in the kinetic term in (6.5), since the DBI action comes from a disk amplitude (and they are in string frame). In fact they do have the right factor in (3.14), but not in (6.5). So we disagree here\dots } 
\be \label{al}
\alpha\equiv \frac{2\pi g_{s}}{v^{s}_{2}}=\frac{2\pi \sqrt{g_{s}}}{v_{2}}\,,
\ee
where now $v_{2}$ is in Einstein frame, and $f$ given by \ref{f}. 


\subsection{D5-brane Phenomenology}

Again the phenomenology is dictated by the size of $\xi$. Focusing on CMB scales we find
\be 
\xi\equiv \frac{\sqrt{2 \epsilon} \alpha}{2f} =\frac{4 \pi^{2}}{\phi \sqrt{2}}\frac{1}{v_{2}}\sqrt{\frac{\V_{E}}{c_{\alpha -- }v^{\alpha}}}\simeq 2.8 \frac{1}{v_{2}}\sqrt{\frac{\V_{E}}{c_{\alpha -- }v^{\alpha}}}\,.
\ee
For the toy model of an isotropic compactification the powers of the volume cancel exactly leaving $\xi$ remarkably close to but slightly above the bound imposed by non-Gaussianity. This means that if a D5-brane is present in the same cycle as the axion, the phenomenology discussed in this paper is a natural outcome. A very explicit construction is needed to evaluate the volume factors which could easily move $\xi$ above or below its experimental bound. 

Another way of discussing the result \eqref{al} is by looking at figure \ref{Fig:param}. A natural size of $\alpha$ is of order $.1$ to $.01$ which makes the inverse decay effect of comparable order as the one induced by oscillations (whose overall size of course depends on the independent parameter $b$).


\section{Conclusions}\label{sec:conclusions}


In this paper, we have studied a very simple and natural class of inflation models wherein $\varphi$ is a pseudo-scalar axion.  We have found that this construction leads to a surprisingly rich array of \emph{correlated} observables for both scalar and tensor cosmological perturbations on CMB scales and also at much smaller scales.  For natural values of the axion decay constant -- roughly $f \sim 10^{-2} M_p$ -- these signals may be detectable in the not-too-distant future.

On CMB/LSS scales, we have argued that a detectable non-Gaussian signature arises naturally.  Non-Gaussianity in axion inflation can feature equilateral and resonant type bispectra or a superposition of the two.  We have reviewed these effects using the in-in formalism and generalized previous results to allow for more general classes of gauge field couplings (which explicitly break the underlying shift symmetry).  Since most axion models are of the large field type, we also expect a large tensor-to-scalar ratio.

We have found that axion inflation may lead also to detectable gravitational wave signatures at Advanced LIGO/VIRGO scales as realized in \cite{GWlorenzo}. We extended that analysis by including the backreaction effects on the late time inflaton dynamics that necessarily arise for couplings leading to observational effects.  In particular, we have noticed that such effects can increase the number of e-foldings by $\sim 10$, which has important consequences for observables. Specifically, it increases the values of $1-n_s$ and $r$ at CMB scales, and of $\Omega_{GW} h^2$ at interferometer scales. The gravity wave signal is correlated with the large-scale non-Gaussianity and, if both effects were detected, it would provide a remarkable confirmation of this scenario.  

We have also noticed that the \emph{same} pseudo-scalar coupling which leads to equilateral non-Gaussianity and small-scale gravitational waves will also provide a natural decay channel for the inflaton.  Indeed, we have seen that reheating is likely to be extremely efficient in the observationally interesting region of parameter space.

Most of our phenomenology relies on the pseudo-scalar coupling $\varphi F\tilde{F}$, which is constrained only by naturalness at the effective field theory level.  In order to address UV sensitivity, we have computed this coupling in specific string theory constructions of axion monodromy.  We have found that such couplings \emph{must} be present to gauge fields living on the NS5 brane.  However, the coupling is large enough to lead to interesting phenomenology only when the moduli stabilization is such that $g_{s}C_{0}\gtrsim 7$, which can be achieved roughly at the cost of a $.5\%$ fine tuning.  
For completeness we noticed that, remarkably, the coupling to a putative D5-brane gauge fields would naturally take the value currently constrained by observations.

Before concluding we would like to mention a few interesting directions for future research. It would be interesting to:
\begin{itemize}
\item Extend to smaller scales the study of the  phenomenological limits due to inverse decay. The running of the spectrum is  larger than for the vacuum contribution and could receive bounds competitive with those from CMB  non-Gaussianity from the inclusion of high $l$ data of experiments like ACT  \cite{Hlozek:2011pc} or SPT \cite{Keisler:2011aw}. Stronger bounds could also emerge by combining 
non-gaussianity at both CMB and LSS scales, taking into account the specific evolution of  of $\xi$ (and of $f_{NL}$) betweeen the two scales.
\item Find the observational bound on resonant non-Gaussianity.
\item Study observational constraints on scales in between CMB and interferometers. Although probes of primordial perturbations at these scales might be less sensitive, and more model dependent (for instance, interesting limits can be obtained if one assumes that the dark matter consists on neutralinos; see \cite{Bringmann:2011ut} for a summary of limits at various scales) the signal is predicted to grow monotonically as one approaches smaller scales (still remaining perturbative).
\item Forecast the constraints that various interferometers could put on the parameters of the model, in case of a detection.
\item Provide a detailed account of (p)reheating in string theory axion monodromy, along the lines of \cite{warped,Kofman:2005yz,Dufaux:2008br,multi,modular}.
\item Study in detail the mechanism that keeps perturbation theory under control and determine how large the tensor-to-scalar ratio can become in the regime of strong backreaction.
\end{itemize}
We hope to address some of these questions in future research. 


\section*{Acknowledgments}

We are grateful to C.~Contaldi, A.~Erickcek, R.~Flauger, R.~Hlozek, L.~McAllister, S.~McWilliams, M.~Nolta, S.~Shandera, L.~Sorbo, T.~Wrase and M.~Zaldarriaga for interesting correspondence and discussions. NB and MP are supported in part by DOE grant DE-FG0294ER-40823 at UMN.


\appendix

\section*{Appendix A: Production of Gauge Fluctuations}
\label{app:gaugeprodn}

In this appendix we study the tachyonic production of gauge field fluctuations in the theory (\ref{Lgen}).  This analysis generalizes the previous work \cite{lorenzo,ai,ai2} to account for generic coupling functions $B(\varphi)$, $C(\varphi)$.  We assume that the slow variation parameters (\ref{SRB}) and (\ref{SRC}) are small.  

Working in Coulomb gauge, the equation of motion of the gauge field in the background of the slowly-rolling inflaton condensate takes the form
\begin{equation}
\label{eom}
  A_i'' + \frac{\phi'}{B}\frac{\partial B}{\partial\varphi} A_i' - \grad^2 A_i 
  - \frac{\phi'}{B}\frac{\partial C}{\partial\varphi}(\grad \times \vec{A})_i = 0 \, .
\end{equation}
We introduce the re-scaled field
\begin{equation}
 \tilde{A}_i(t,{\bf x}) \equiv \sqrt{B(\phi(t))} \, A_i(t,{\bf x}) \, ,
\end{equation}
and decompose $\tilde{A}_i$ into circular polarization modes, as in equation (\ref{Adecomp}).  The equation of motion is:
\begin{equation}
\label{genmode}
  \left[ \frac{\partial^2}{\partial\tau^2} + k^2 + M_A^2(\tau) \pm \frac{2 k\xi}{\tau} \right] {\tilde A}_{\pm}(k,\tau) = 0 \, ,
\end{equation}
where
\begin{eqnarray}
  M^2_A(\tau) &\cong& - \frac{\sqrt{\epsilon_\phi \epsilon_B}}{\tau^2} \, {\rm sign} \left( B' \dot{\phi} \right) \label{Meffsquared} \, , \\
 \xi &\equiv& \sqrt{\frac{\epsilon_\phi}{2}} M_p \frac{C'}{B} \mathrm{sign}\left(\dot{\phi}\right)\, . \label{xigen}
\end{eqnarray}
where $\epsilon_\phi \equiv \frac{\dot{\phi}^2}{2H^2 M_p^2}$, which may differ from $\epsilon_V$ (see Eqn.~\ref{SR}) or $-\dot{H}/H^2$ in the regime of strong backreaction.

In equation (\ref{Meffsquared}) we work to leading order in slow-variation parameters.  In the following, for brevity, we assume that $\dot{\phi} C' > 0$ and $B' \dot{\phi}  > 0$.  One can show that
\begin{equation}
 \frac{\dot{\xi}}{ H \xi} \cong \frac{\ddot{\phi}_0}{H\dot{\phi}_0} + \epsilon_\phi  + 2 \left( \sqrt{ \epsilon_\phi \, \epsilon_C} - \sqrt{\epsilon_\phi \, \epsilon_B} \right) \, .
 \end{equation}
This shows that, when the slow-variation constraints are respected, we can treat $\xi$ as a constant.  In this case, equation (\ref{genmode}) has the form of Whittaker's equation
\begin{equation}
\label{white}
 \left[ \frac{d^2}{dz^2} - \frac{1}{4} + \frac{\lambda}{z} + \frac{1/4-\mu^2}{z^2} \right] W_{\lambda,\mu}(z) = 0 \, .
\end{equation}
The gauge mode solutions may therefore be expressed in terms of Whittaker's function.  The correctly normalized solution is
\begin{eqnarray}
  && A_+(k,\tau) = \frac{1}{\sqrt{2k \, B_*}}\, e^{\pi \xi / 2} \, W_{-\kappa,\mu}(+2ik\tau) \, , \label{Amode2} \\
  && \kappa \equiv i\xi, \hspace{5mm} \, \mu \cong \frac{1}{2} + \sqrt{\epsilon_\phi\epsilon_B} \, .
\end{eqnarray}
where $B_\star$ is $B(\phi(t))$ evaluated at horizon crossing for cosmologically interesting modes.  Equation (\ref{Amode2}) can be put in a more illuminating form by using the asymptotic expansion
\begin{equation}
\label{whittaker}
  W_{-\kappa,\mu}(z) \cong \left(\frac{z}{4\kappa}\right)^{1/4} \kappa^{-\kappa} e^{\kappa} e^{-2\sqrt{\kappa z}} \, ,
\end{equation}
 which is valid for large  $|\kappa|$ and when $\mathrm{Im}(\kappa) > 0$.  Equation (\ref{whittaker}) leads immediately to the following limiting behaviour for (\ref{Amode2}):
\begin{equation}
\label{Amode3}
   A_+(\tau,k) \cong \frac{1}{\sqrt{2k \, B_*}}\left(\frac{k}{2\xi a H}\right)^{1/4} e^{\pi \xi - 2\sqrt{2\xi k / (aH)}} \, .
\end{equation}
Formally, this expression is only valid for $\xi \rightarrow \infty$.  However, one may verify numerically that it provide a good fit even for $\xi \sim 2-3$, at least for modes with $(8\xi)^{-1} \lsim -k\tau \lsim 2\xi$ which account for most of the power in produced fluctuations.  

Equation (\ref{Amode3}) differs slightly from the result that was obtained in \cite{ai,ai2} for the case $B=1$ and $C\propto\varphi$.  The key ``new'' feature encoded in this equation is the fact that, if $B\ll 1$ during inflation, then the value of $\xi$ is effectively increased.  In this case, one can find a much greater production of gauge quanta, since the amplitude of $A_i$ is exponentially sensitive to $\xi$. 

Notice that equation (\ref{Amode3}) will fail to adequately describe the dynamics on sufficiently large scales.  This expression results from neglecting the last term in square braces of (\ref{white}), an assumption that breaks down for $-k\tau < \mathcal{O}(\epsilon /\xi)$.  From (\ref{eom}), we can see that the ``new'' interaction associated with $B(\varphi)$ actually dominates the dynamics of $A_\mu$ in the limit $-k\tau \rightarrow 0$.  Hence, we should verify that this interaction does not lead to any interesting particle production on super-horizon scales, that would not be accounted for by the asymptotic expansion (\ref{Amode3}).  Neglecting gradients, the equation of motion (\ref{eom}) becomes
\begin{equation}
 \ddot{A}_i + \left[H + \frac{1}{B}\frac{dB}{dt}\right]\dot{A}_i \cong 0 \, .
\end{equation}
Suppressing the vector index, the solution is
\begin{equation}
 A = A_\star + \dot{A}_\star \int dt \frac{a_\star}{a} \frac{B_\star}{B} \, .
\end{equation}
Our assumption about slow-variation of $B(t)$ means that $(a B)^{-1}$ should be a decreasing function of $t$ and the integral will rapidly converge.  Assuming that $B$ varies slowly enough to pull it out of the integration we have
\begin{equation}
 A \cong A_\star + \frac{B_\star}{B} \frac{\dot{A}_\star}{H} \left(1 - \frac{a_\star}{a}\right) \rightarrow \mathrm{const} \, .
\end{equation}
This shows explicitly that the coupling $B(\varphi)$ does not lead to any interesting particle production on super-horizon scales.



\section*{Appendix B: Effective Action for the Perturbations}
\label{app:effectiveaction}

In this Appendix, we derive the leading interactions terms in the effective action for the curvature perturbation.  Following \cite{ai,ai2} and \cite{leblond} we neglect metric perturbations.  The action for $\delta\varphi$, including terms up to third order in fluctuations, is given by
\begin{eqnarray}
 S &=& \int dt d^x x a^3 \left[  \frac{1}{2}(\delta\dot{\varphi})^2 - \frac{1}{2a^2}(\vec{\nabla}\delta\varphi)^2  
         - \frac{1}{2}\left[m^2 - \frac{\Lambda^4}{f^2}\cos\left(\frac{\phi_0(t)}{f}\right)\right] (\delta\varphi)^2 
         - \frac{1}{6} \frac{\Lambda^4}{f^3}\sin\left(\frac{\phi_0(t)}{f}\right)(\delta\varphi)^3  \right] \nonumber \\
  &+& \int dt d^3x a^3 \left[ - \frac{B'}{4} \delta\varphi F^{\mu\nu}F_{\mu\nu} - \frac{C'}{4} \delta\varphi F^{\mu\nu}\tilde{F}_{\mu\nu}  \right] \label{Lapp}
\end{eqnarray}
Here we have introduced the notation $m^2 \equiv V_{\mathrm{sr}}''$ and we write the zeroth order homogeneous solution $\phi_0(t)$ in the trigonometric functions, since the difference $\phi(t) - \phi_0(t)$ is subleading in the small parameter $b\equiv \frac{\Lambda^4}{f V_{\mathrm{sr}}'}$; see (\ref{phi_bkg}).  Moreover, we have disregarded a contribution proportional to $V_{\mathrm{sr}}'''$, which is usually subleading in a slow roll expansion and is actually vanishing for string theory axion monodromy inflation.

Our goal is to re-write the action (\ref{Lapp}) using
\begin{equation}
\label{zeta_app}
 \zeta(t,{\bf x}) = -\frac{H}{\dot{\phi}} \, \delta\varphi(t,{\bf x}) \hspace{5mm} \Rightarrow \hspace{5mm} \delta\varphi = -\sqrt{2\epsilon} M_p \zeta
\end{equation}
and identify the leading interaction terms.  Let us focus, first, on the quadratic terms on the first line of (\ref{Lapp}).  Using (\ref{zeta_app}) is is straightforward to derive the result
\begin{equation}
\label{S2app_almost}
 S \supset M_p^2 \int dt d^3 x a^3 \epsilon \left[ \dot{\zeta}^2 - \frac{1}{a^2}(\vec{\nabla}\zeta)^2 - \Omega^2\zeta^2 \right]
\end{equation}
where we have defined
\begin{equation}
\label{Omega}
 \Omega^2 \equiv m^2 - \frac{\Lambda^4}{f^2}\cos\left(\frac{\phi_0(t)}{f}\right) 
  -H^2 \left( \frac{\dot{\delta}}{H} + 3\epsilon+3\delta + 2\epsilon^2+3\epsilon\delta+\delta^2\right)
\end{equation}
and introduced the parameter $\delta \equiv \ddot{H} / (2 H \dot{H})$.  In the case without resonance effects (when $b=0$) the last term in (\ref{S2app_almost}) would represent a tiny and unimportant correction.  However, it is important to notice that it is $\dot{\phi}$, and not $\dot{\phi}_0$, appears in (\ref{zeta_app}), hence there is rapid oscillatory time dependence implicit in the parameters $\epsilon$, $\delta$ appearing in $\Omega^2$ which could, in principle, lead to interesting effects.  To extract the oscillatory terms from $\Omega^2$, we expand the slow roll parameters as
\begin{equation}
   \epsilon = \epsilon_0 + \epsilon_1 + \cdots \, , \hspace{5mm} \delta = \delta_0 + \delta_1 + \cdots \label{srexpand}
\end{equation}
The subscript $i$ indicates that a quantity is computed at $i$-th order in $b$.  The leading corrections are given by \cite{pajer,leblond}
\begin{eqnarray}
 \epsilon_1 &=& -3 b \sqrt{2\epsilon_0} \frac{f}{M_p} \cos\left(\frac{\phi_0(t)}{f}\right) \, , \label{e1a}\\
 \delta_1 &=& -3 b \sin\left(\frac{\phi_0(t)}{f}\right) \label{d1a} \, ,
\end{eqnarray}
which shows that $\delta \gg \epsilon$.  Using (\ref{e1a}) and (\ref{d1a}) it is straightforward to check that oscillatory contribution to $\Omega^2$ cancels at leading order.  Hence, we are left with 
\begin{equation}
\label{S2app}
 S_2 \approx M_p^2 \int dt d^3 x a^3 \epsilon \left[ \dot{\zeta}^2 - \frac{1}{a^2}(\vec{\nabla}\zeta)^2  \right] \, ,
\end{equation}
which gives the same linearized equation of motion that was studied in \cite{leblond}
\begin{equation}
 \zeta'' + 2 a H (1+\delta)\zeta' - \vec{\nabla}^2 \zeta \approx 0 \, .
\end{equation}

We take the ``free theory'' to correspond to (\ref{S2app}), neglecting resonance effects (setting $b=0$), and treat all other contributions as perturbative ``interactions''.  That is, we identify the free theory Lagrangian as
\begin{equation}
\label{Lfree_app}
  \mathcal{L}_0 = M_p^2 a^3 \epsilon_0 \left[ \dot{\zeta}^2 - \frac{1}{a^2}(\vec{\nabla}\zeta)^2  \right] \, .
\end{equation}
The $\mathcal{O}(b)$ correction to the quadratic Lagrangian gives
\begin{equation}
 \mathcal{L}_I \supset M_p^2 a^3 \epsilon_1 \left[ \dot{\zeta}^2 - \frac{1}{a^2}(\vec{\nabla}\zeta)^2  \right] \, .
\end{equation}
There is also a cubic interaction arising from the last term on the first line of (\ref{Lapp}).  To leading order in $b$ this gives
\begin{equation}
 \mathcal{L}_I \supset \frac{a^3}{6} (2\epsilon_0)^{3/2} \Lambda^4 \frac{M_p^3}{f^3} \sin\left(\frac{\phi_0(t)}{f}\right) \zeta^3 \, .
\end{equation}

Next, we turn our attention to the gauge field interactions on the second line of (\ref{Lapp}).  At leading order we have
\begin{equation}
\label{Lgauge1app}
 \mathcal{L}_I \supset a^3 B(\phi_0) \left[ \frac{\xi}{2} \zeta F\tilde{F} + \mathrm{sign}(B')\frac{\sqrt{\epsilon \epsilon_B}}{2} \zeta F^2 \right] \, .
\end{equation}
Notice that, when particle production effects are relevant, we always have $\xi \gsim 1 \gg \sqrt{\epsilon\epsilon_B}$.  Thus, the first term in (\ref{Lgauge1app}) always dominates in the observationally interesting regime.  We can further simplify the result by making a field redefinition $A_\mu \rightarrow B^{-1/2} A_\mu$ and neglecting derivatives of $B$, which is justified to leading order in the slow-variation parameters (\ref{SRB}).  Thus, we arrive at
\begin{equation}
 \mathcal{L}_I \supset \frac{\xi}{4} \,\zeta\, \eta^{\mu\nu\alpha\beta} F_{\mu\nu} F_{\alpha\beta}
\end{equation}
where $\epsilon^{\mu\nu\alpha\beta}$ is the Minkowski-space Levi-Civita symbol, with $\eta^{0123} = +1$.

Adding up all the relevant terms, we arrive at the following result interaction Lagrangian
\begin{equation}
 \mathcal{L}_I \cong M_p^2 a^3 \epsilon_1 \left[ \dot{\zeta}^2 - \frac{1}{a^2}(\vec{\nabla}\zeta)^2  \right] 
                + \frac{a^3}{6} (2\epsilon_0)^{3/2} \Lambda^4 \frac{M_p^3}{f^3}\sin\left(\frac{\phi_0(t)}{f}\right)\, \zeta^3 
                  + \frac{\xi}{4} \,\zeta \,\eta^{\mu\nu\alpha\beta} F_{\mu\nu} F_{\alpha\beta} \label{LIfinalapp}
\end{equation}
Equation (\ref{LIfinalapp}) is the main result of this Appendix.



\section*{Appendix C: Inverse Decay Effects with the In-In Formalism}
\label{app:inv.dec}

In this Appendix we study inverse decay effects using the in-in formalism, showing that this reproduces exactly the results which were derived in \cite{ai,ai2} using a different method.  The relevant term in the interaction Lagrangian (\ref{LIfinalapp}) is
\begin{equation}
\label{LIinvdec}
 \mathcal{L}_I^{\mathrm{inv.dec}} = + a^3 \frac{\xi}{2} \zeta F_{\mu\nu} \tilde{F}^{\mu\nu} \, .
\end{equation}
This coincides with the interaction that was studied in \cite{ai,ai2} (only the meaning of the constant $\xi$ is changed due to the presence of non-trivial coupling functions $B(\varphi)$ and $C(\varphi)$).  The interaction Hamiltonian is given by $H_I = -\int d^3x \mathcal{L}_I$ and may be written in the form
\begin{equation}
\label{HIinvdec}
 H_I^{\mathrm{inv.dec}}(t) = \sqrt{2\epsilon_0} M_p \, \int d^3 q \zeta_{\bf -q}(t) J_{\bf q}(t) \, ,
\end{equation}
to leading order.  Here the ``source'' term is defined by
\begin{equation}
 J_{\bf q}(t) \equiv -\frac{a^3}{2\sqrt{2\epsilon_0}} \frac{\xi}{M_p} \int \frac{d^3x}{(2\pi)^{3/2}} \, e^{-i{\bf k}\cdot {\bf x}}\,
  \left[ F^{\mu\nu}\tilde{F}_{\mu\nu} - \langle F^{\mu\nu} \tilde{F}_{\mu\nu} \rangle \right] \, .
\end{equation}
Notice that $\langle J_{\bf q}(t) \rangle = 0$ by construction.  

The quantity $J_{\bf q}$ has the same meaning as in \cite{ai,ai2}: it is the source term in the equation of motion for $\delta\varphi$ that corresponds to the production of inflaton fluctuations by inverse decay.  Defining the canonical field variable
\begin{equation}
\label{Qdef}
 Q_{\bf k}(t) \equiv -\sqrt{2\epsilon_0} M_p a \, \zeta_{\bf k} \, ,
\end{equation}
the equation of motion becomes
\begin{equation}
\label{IDeom}
 \left[ \partial_\tau^2 + k^2 - \frac{a''}{a}  \right] Q_{\bf k}(t) = J_{\bf k}(t) + \cdots
\end{equation}
where $\cdots$ denotes additional corrections, arising from resonance effects and self-interactions, which will not concern us here.  We are interested in the particular solution of (\ref{IDeom}), which corresponds to the inflaton fluctuations generated by inverse decay and may be highly non-Gaussian.  Working at the level of the equation of motion, it is straightforward to see that the inverse decay contribution to the $n$-point correlation functions of the curvature perturbation may be written as
\begin{equation}
\label{IDeomNpt}
 \left.\langle\zeta_{\bf k_1}\zeta_{\bf k_2} \cdots \zeta_{\bf k_n}\rangle \right|_{\mathrm{inv.dec}} 
  = \left(-\frac{1}{\sqrt{2\epsilon_0} M_p}\right)^n \left[ \prod_{i=1}^n \int_{-\infty}^\tau d\tau_i \frac{G_{k_i}(\tau,\tau_i)}{a(\tau)} \right]
  \times \langle J_{\bf k_1}(\tau_1) J_{\bf k_2}(\tau_2) \cdots J_{\bf k_n}(\tau_n) \rangle
\end{equation}
where $G_k(\tau,\tau')$ is the retarded Green function for equation (\ref{IDeom}); see \cite{ai2} for details.  Here we show that the same result may be derived also from the in-in formalism.

Focusing on the interaction term (\ref{HIinvdec}) the in-in formula (\ref{in-in}) is
\begin{eqnarray}
 \langle \zeta_{\bf k_1} \zeta_{\bf k_2} \cdots \zeta_{\bf k_n} (\tau) \rangle &=& 
 \sum_{N=0}^{\infty} (-i)^N \int^t dt_1 \int^{t_1} dt_2\cdots \int^{t_{N-1}} dt_N 
   \nonumber \\
 && \times \, \langle\left[\left[\left[ \zeta_{\bf k_1} \zeta_{\bf k_2}\cdots \zeta_{\bf k_n}(\tau), H_I^{\mathrm{inv.dec}}(t_1)   \right],  H_I^{\mathrm{inv.dec}}(t_2)\right] \cdots, H_I^{\mathrm{inv.dec}}(t_N) \right] \rangle \nonumber \\
&=&  \sum_{N=0}^{\infty} \left(-i \sqrt{2\epsilon_0} M_p \right)^N \left[ \prod_{i=1}^N \int^{t_{i-1}} dt_i \int d^3 q_i  \right] \nonumber \\
&&  \times \, \langle\left[\left[\left[ \zeta_{\bf k_1} \zeta_{\bf k_2}\cdots \zeta_{\bf k_n}(\tau), \zeta_{\bf -q_1}(t_1) J_{\bf q_1}(t_1)  \right],  \zeta_{\bf -q_2}(t_2) J_{\bf q_2}(t_2)\right] \cdots, \zeta_{\bf -q_N}(t_N) J_{\bf q_N}(t_N) \right] \rangle \, , \label{in-in-invdec}
\end{eqnarray}
where $t_0 \equiv t$ in the product over integrals.  A crucial simplification arises from noting that the ``source'' terms $J_{q_i}(\tau_i)$ may be treated as commuting variables, to a very good approximation.  This follows from the discussion in Subsection \ref{subsec:gaugeprodn}; see equation (\ref{Acommute}) in particular.  This allows us to pull the factors of $J_{q_i}(\tau_i)$ out of the nested commutator in (\ref{in-in-invdec}).  Thus, we arrive at a simplified formula
\begin{eqnarray}
 \langle \zeta_{\bf k_1} \zeta_{\bf k_2} \cdots \zeta_{\bf k_n} (\tau) \rangle
&=& \sum_{N=0}^{\infty} \left(-i\sqrt{2\epsilon_0} M_p \right)^N \left[ \prod_{i=1}^N \int^{t_{i-1}} dt_i \int d^3 q_i J_{\bf q_i}(\tau_i)  \right] \nonumber \\
&&  \times \, \langle\left[\left[\left[ \zeta_{\bf k_1} \zeta_{\bf k_2}\cdots \zeta_{\bf k_n}(\tau), \zeta_{\bf -q_1}(t_1) \right],  \zeta_{\bf -q_2}(t_2) \right] \cdots, \zeta_{\bf -q_N}(t_N)  \right] \rangle \, . \label{in-in-invdec-final}
\end{eqnarray}

To evaluate the nested commutators on the last line of (\ref{in-in-invdec-final}) we can use the formula
\begin{eqnarray}
 \left[\zeta_{\bf k_1}(\tau_1), \zeta_{\bf k_2}(\tau_2)\right] &\cong& \frac{1}{2\epsilon_0} \frac{1}{M_p^2} \left[Q_{\bf k_1}(\tau_1), Q_{\bf k_2}(\tau_2)\right] \nonumber \\
 &=& \frac{-i}{2\epsilon_0 M_p^2} \frac{G_{k_1}(\tau_1,\tau_2)}{a(\tau_1)a(\tau_2)} \delta^{(3)}\left({\bf k_1}+{\bf k_2}\right) \, , \label{commutator_identity}
\end{eqnarray}
where the second equality is valid only for $\tau_1\geq \tau_2$.

Let us first consider the 2-point function; equation (\ref{in-in-invdec-final}) with $n=2$.  The $N=0$ term is the summation just gives the usual (nearly) scale-invariant power spectrum from the quantum vacuum fluctuations.  Inverse decay effects, on the other hand, are encoded in the $N\geq 1$ terms.  Using (\ref{commutator_identity}) it is straightforward to verify that only the $N=2$ contribution to the nested commutators gives a non-vanishing contribution.  Explicit evaluation gives
\begin{equation}
 \left. \langle \zeta_{\bf k_1} \zeta_{\bf k_2}(\tau) \rangle\right|_{\mathrm{inv.dec}} \approx
 \left(\frac{-1}{\sqrt{2\epsilon_0}M_p}\right)^2 \int_{-\infty}^{\tau} d\tau_1 d\tau_2 \, G_{k_1}(\tau,\tau_1) G_{k_2}(\tau,\tau_2) \, 
  \langle J_{\bf k_1}(\tau_1) J_{\bf k_2}(\tau_2) \rangle
\end{equation}
in precise agreement with (\ref{IDeomNpt}).

Next, we turn our attention to the 3-point function.  This time the $N=0$ term is trivial, since the free theory modes are gaussian.  It may be readily verified that only the $N=3$ term in the summation (\ref{in-in-invdec-final}) survives; all other commutators are zero.  Hence, we find
\begin{equation}
 \left. \langle \zeta_{\bf k_1} \zeta_{\bf k_2}\zeta_{\bf k_3}(\tau) \rangle\right|_{\mathrm{inv.dec}} \approx
 \left(\frac{-1}{\sqrt{2\epsilon_0}M_p}\right)^3 \int_{-\infty}^{\tau} d\tau_1 d\tau_2 d\tau_3\,  G_{k_1}(\tau,\tau_1) G_{k_2}(\tau,\tau_2) G_{k_3}(\tau,\tau_3) \,
  \langle J_{\bf k_1}(\tau_1) J_{\bf k_2}(\tau_2) J_{\bf k_3}(\tau_3) \rangle
\end{equation}
which, again, agrees with (\ref{IDeomNpt}).The analysis of this Appendix proves that the formalism developed in \cite{ai,ai2} is equivalent to the in-in method.

\section*{Appendix D: Resonance Effects with the In-In Formalism}
\label{app:res}

In this Appendix we study resonance effects \cite{chen2,pajer,leblond} on the 2-point and 3-point function, using the in-in formalism.  These effects are generated by the first two terms in the interaction Lagrangian (\ref{LIfinalapp}):
\begin{equation}
 \mathcal{L}_I \supset M_p^2 a^3 \epsilon_1 \left[ \dot{\zeta}^2 - \frac{1}{a^2}(\vec{\nabla}\zeta)^2  \right] 
                + \frac{a^3}{6} (2\epsilon_0)^{3/2} \Lambda^4 \frac{M_p^3}{f^3}\sin\left(\frac{\phi_0(t)}{f}\right) \zeta^3 \, .
\end{equation} 

Let us first consider the 2-point function.  The relevant term in the interaction Hamiltonian is
\begin{equation}
 H_I(t) \supset -\int d^3 x a^3 \epsilon_1 \left[ \dot{\zeta}^2 - \frac{1}{a^2}(\vec{\nabla}\zeta)^2  \right] \, . 
\end{equation}
The $N=1$ term in the in-in formula (\ref{in-in}) gives
\begin{equation}
 \langle \zeta_{\bf k_1} \zeta_{\bf k_2}(\tau) \rangle \supset i M_p^2 \int_0^t dt_1 d^3x \epsilon_1(t)  
\Bigg \langle \left[ \zeta_{\bf k_1}\zeta_{\bf k_2}(\tau), (\dot{\zeta}(t_1,{\bf x}))^2 - \frac{1}{a^2}(\vec{\nabla}\zeta(t_1,{\bf x}))^2   \right] \Bigg \rangle \, .
\label{in-in-res-2}
\end{equation}
Recall that the oscillatory time dependence of $\epsilon_1$ is given by (\ref{e1a}).  It is straightforward (but non-trivial) to evaluate (\ref{in-in-res-2}).  We find
\begin{equation}
 \langle \zeta_{\bf k_1} \zeta_{\bf k_2}(\tau) \rangle \supset \frac{2 H^4}{\dot{\phi}_0^2 k^3} \, \delta^{(3)}\left({\bf k_1}+{\bf k_2}\right) \,
 \int_{\infty}^x dx' \left[\frac{3bf}{\sqrt{2\epsilon_0}}\right] \sin(2x') \cos\left(\frac{\phi_0}{f}\right) \, ,
\end{equation}
where $x \equiv -k_1\tau$.  We can write the homogeneous solution as $\phi_0 = \phi_k + \sqrt{2\epsilon_0} M_p \ln x$; see \cite{pajer}.  Following \cite{pajer}, we introduce the quantity
\begin{equation}
 c^{(-)} \equiv -3 b i \frac{f}{\sqrt{2\epsilon_0}} \int_{\infty}^x dx' e^{2ix'} \cos\left(\frac{\phi_k}{f} + \frac{\sqrt{2\epsilon_0} M_p}{f} \ln x' \right)
\end{equation}
In this case the total 2-point function, accounting also for the $N=0$ term in (\ref{in-in}), can be written as
\begin{equation}
 \langle \zeta_{\bf k_1} \zeta_{\bf k_2}(\tau) \rangle \equiv \frac{2\pi^2}{k_1^3} P_\zeta(k_1) \delta^{(3)}\left({\bf k_1}+{\bf k_2}\right)
\end{equation}
where
\begin{equation}
 P_\zeta(k) = \mathcal{P} \left( 1 + 2 \mathrm{Re} \left[ c^{(-)} \right] \right)
\end{equation}
This coincides with the result of \cite{pajer} that was obtained using a different method.

Next, we turn our attention to the 3-point function.  The relevant term in the interaction Hamiltonian is
\begin{equation}
 H_I(t) \supset - \int d^3x \frac{a^3}{6} (2\epsilon_0)^{3/2} \Lambda^4 \frac{M_p^3}{f^3}\sin\left(\frac{\phi_0(t)}{f}\right) \zeta^3
\end{equation}
This interaction was taken into account using the in-in method in \cite{leblond}, where it was shown that the bispectrum agrees with \cite{monodromy2}.



\section*{Appendix E: S-duality}\label{a:S}

S-duality is part of the $SL(2,\mathbb{R})$ symmetry of the classical action Type IIB (broken to $SL(2,\mathbb{Z})$ at the quantum level). The 10D Einstein frame metric $G_{E}$ is related to the string frame metric $G_{s}$ by $G_{E,MN}=e^{-\phi/2}G_{s,MN}$. It is $G_{E}$ that is invariant under $SL(2,\mathbb{R})$ and \textit{not} $G_{s}$. In the 10D Einstein frame, $SL(2,\mathbb{R})$ acts as \cite{Polchinski:1998rr}
\be 
\tau'=\frac{a\tau+b}{c\tau+d}\,,\quad 
\left(\begin{array}{c} H_{3}' \\ F_{3}' \end{array}\right)
=\left(
\begin{array}{cc}
d&c\\
b&a
\end{array}
\right) \left(\begin{array}{c} H_{3} \\ F_{3} \end{array}\right)\,,
\ee
with $ad-bc=1$, where $\tau\equiv C_{0}+i e^{-\phi}$ and all the other fields being invariant. S-duality corresponds then to the choice $a=d=0$ and $b=-c=\pm1$, leading to the transformation $\tau'=-1/\tau$ or
\be
C_{0}'=-\frac{C_{0}} {C_{0}^{2}+e^{-2\phi}} \,,\quad e^{-\phi'}=\frac{e^{-\phi}} {C_{0}^{2}+e^{-2\phi}}\,.
\ee
The D-brane action in Einstein frame is \cite{Polchinski:1998rq}
\be 
S_{DBI}&=&-\mu_{5}\int d^{6}\xi e^{-\phi} \sqrt{-{\rm det} (e^{\phi/2}G_{E}^{\rm ind}+B^{\rm ind}+2\pi \alpha' F)}\,,
\ee
Then by S-duality one finds the NS5-brane action \cite{Eyras:1998hn}
\be
S_{DBI}&=& -\mu_{5}\int d^{6}\xi e^{-\phi}|\tau| \sqrt{-{\rm det} \left[e^{\phi/2}G_{E}^{\rm ind}-|\tau|^{-1}\left(C^{\rm ind}+2\pi \alpha' \tilde F\right)\right]}\\
&=&-\mu_{5}\int d^{6}\xi \frac{\sqrt{1+e^{2\phi}C_{0}^{2}}} {e^{2\phi}} \\
&&\quad\sqrt{-{\rm det} \left[e^{\phi/2}G_{E}^{\rm ind}-\frac{e^{\phi}}{\sqrt{1+e^{2\phi}C_{0}^{2}}}\left(C^{\rm ind}+2\pi \alpha' \tilde F\right)\right]}\,,
\ee
where $\tilde F=d\tilde A_{1}$ is the field strength of the vector field on the worldvolume of the NS5-brane. If we assume $C_{0}=0$ and compactify to four-dimensions we find\footnote{Note that here we have made a different choice of the four-dimensional Einstein frame with respect to \cite{monodromy}. There the four-dimensional metric is the string metric and the extra factors of $g_{s}$ are reabsorbed in the definition of the Planck constant (which is hence also different from what we have here). The advantage of the present choice is that the $SL(2,\mathbb{Z})$ transformations are simpler in the ten-dimensional Einstein frame. On the other hand, one has to remember to go to Einstein frame when using the DBI action!}
\be
S\supset -\mu_{5}\alpha'\int d^{4}\xi \sqrt{-g_{E,4}}\sqrt{l_{E}^{4}e^{-\phi}- c^{2}}\,,
\ee
where $l_{E}^{2}\al$ is the volume of the two-cycle measured with the Einstein metric. One can check that the tension of the NS5-brane (in string frame) has an extra factor of $g_{s}^{-1}$ with respect to the tension of a D5-brane as expected since the two 5-branes are related by S-duality.


\end{document}